\pdfoutput=1

\documentclass[journal,twocolumn,12pt]{IEEEtran}

\usepackage{amsmath}
\usepackage{amssymb}
\usepackage{amsfonts}
\usepackage{mathtools}

\usepackage{graphicx}
\graphicspath{{./}}
\usepackage{placeins}

\usepackage[caption=false,font=footnotesize]{subfig}

\usepackage{booktabs}
\usepackage{multirow}
\usepackage{array}
\usepackage{threeparttable}
\usepackage{siunitx}
\newcolumntype{L}[1]{>{\raggedright\arraybackslash}m{#1}}

\usepackage{algorithm}
\usepackage{algorithmic}

\usepackage{textcomp}
\usepackage{xcolor}
\usepackage{balance}

\usepackage{url}
\usepackage[colorlinks=true,
            linkcolor=black,
            citecolor=black,
            urlcolor=blue,
            breaklinks=true,
            bookmarks=false,
            pdfborder={0 0 0}] {hyperref}

\usepackage[capitalise,nameinlink]{cleveref}
\crefname{equation}{}{}
\Crefname{equation}{}{}
\crefname{figure}{Fig.}{Figs.}
\Crefname{figure}{Fig.}{Figs.}
\crefname{table}{Table}{Tables}
\crefname{section}{Section}{Sections}

\newcommand{\etal}{\textit{et al.}}

\begin{document}

\title{The U.S. Interconnection Queue System: Cascading Vulnerability Analysis and A Resilience Engineering Framework}

\author{Zahra~Heidari\\[2pt]
    \normalsize Department of Systems Engineering\\
    \normalsize Colorado State University, Fort Collins, Co, USA\\
    \normalsize \texttt{zahra.heidari@colostate.edu}

    \thanks{Department of Systems Engineering, Colorado State University, Fort Collins, CO, USA (e-mail: zahra.heidari@colostate.edu).}
}

\maketitle

\begin{abstract}

As of 2025, the U.S. interconnection queues, the primary gateway to grid access for new energy generation and storage, contain roughly \(8\text{,}200\) projects totaling \(2\text{,}061\) GW of capacity seeking grid interconnection. Only \(13\%\) of the capacity queued between 2000 and 2020 (\(19\%\) by project count) has reached operation. We argue that the queue architecture is vulnerable to self-reinforcing propagation of project withdrawals arising from restudy and cost reallocation, independent of study processing delays. Adapting failure contagion models from financial and interdependent infrastructure networks to interconnection queue mechanisms, we formulate the queue process as a complex adaptive system, and test it with two analyses: a statistical test of withdrawal temporal clustering and co-withdrawal across seven major ISO/RTOs, and a stylized network contagion simulation of cost-sharing interdependencies with ''circuit-breaker'' interventions. Using project-level data from Lawrence Berkeley National Laboratory (LBNL) through 2025, we find significant temporal clustering (dispersion indices \(5.0\text{--}102.4\); all \(p < 0.001\)). We also identify \(39\) monthly withdrawal bursts with the largest cluster-dated, reaching up to \(67.4\times\) the regional mean. The withdrawal timing is concentrated among projects within technology categories in all seven regions (\(1\text{,}000\)-permutation technology-label test; \(z = 2.58\text{--}4.78\), \(p \leq 0.005\)). The co-withdrawal behavior within cohorts remains statistically significant in four of seven regions (\(0.5\text{--}1.8\) percentage points). The model yields \(1.0\text{--}1.5\times\) the net cascade amplification at low-to-moderate connectivity (average network degree \(k = 3\text{--}10\)). At high connectivity (\(k = 20\)), a phase-transition-like shift to systemic instability occurs, exceeding \(8.7\times\) the net amplification at a \(10\%\) shock (\(97.7\%\) cascade size); a boundary condition of the stylized network that disappears under pro-rata cost redistribution. Circuit-breaker interventions that cap per-neighbor cost reallocation reduce modeled cascade magnitude by up to \(20\%\) at moderate connectivity, with a \(44\%\) lower mean per-neighbor cost transfer, relative to the baseline. These findings characterize the U.S. interconnection queue as critical infrastructure with cascading vulnerability whose severity is conditional on cost allocation design.
\end{abstract}


\IEEEpeerreviewmaketitle




\section{Introduction}
\label{sec:introduction}

\subsection{Scale and Attrition Statistics}
\label{subsec:scale_and_attrition_statistics}
\IEEEPARstart{T}{}he U.S. interconnection queue has been the primary bottleneck for new generation and storage seeking grid interconnection, and existing federal reforms have not yet fully addressed it \cite{ferc2023}. This grid-access bottleneck constrains the pace of energy transition and the deployment of clean energy capacity to achieve decarbonization of the electricity sector. The principal clean energy technologies needed to decarbonize the sector are commercially mature and represent a dominant share of the capacity queued for deployment on large scales \cite{rand2026}. However, the rate at which the planned capacity accesses the grid and reaches commercial operation remains low. 

As of year-end 2025, approximately \(8\text{,}200\) projects reported by the Lawrence Berkeley National Laboratory (LBNL) are seeking grid interconnection across seven major U.S. independent system operators and regional transmission organizations (ISO/RTOs) and more than 50 transmission system operators, and that account for \(1\text{,}312\) GW of generation and \(749\) GW of storage capacity \cite{rand2026}. A direct analysis of the LBNL Queued Up 2026 public data identifies \(8\text{,}513\) project records flagged active, totaling approximately \(1\text{,}865\) GW, and the reported difference arises because the LBNL published capacity accounts for imputed storage capacity for hybrid and co-located projects. At roughly \(2\text{,}061\) GW in total, the queued capacity exceeds the total installed U.S. generating capacity (approximately \(1\text{,}280\) GW; \cite{eia2025}), a figure that itself  has required decades to accumulate. Solar comprises a significant share of queued generation capacity at \(773\) GW and storage at \(749\) GW, followed by wind at \(220\) GW. Natural gas capacity increased \(86\%\) year-over-year to \(253\) GW, which is partly attributed to the growing demand for development of hyperscale data centers \cite{rand2026}. Zero-carbon capacity (wind, solar, nuclear, geothermal, and hydro) accounts for approximately \(77\%\) of the queued generation capacity, compared to \(19\%\) for natural gas.

According to the latest LBNL data, only \(13\%\) of the planned capacity (\(19\%\) by project count) with interconnection requests submitted between 2000 and 2020 reached commercial operation by the end of 2025; approximately \(76\%\) were withdrawn and \(10\%\) remained active in the queue system \cite{rand2026}. The median time between a request and commercial operation has more than doubled from less than two years for projects that reached commercial operation between 2000 and 2007 to \(5.2\) years for those that reach operation between 2024 and 2025 \cite{rand2026}. These estimates are consistent with the project attrition rates reported in \cite{gorman2025grid} that withdrawal rates remain roughly \(80\%\) by project count, and interconnection costs have increased over the same period. The LBNL Queued Up 2026 Edition dataset yields roughly \(71\%\) by project count for the \(2000\text{--}2020\) cohort (Table~\ref{tab:table_1}). 

\begin{table*}[!t]
  \renewcommand{\arraystretch}{1.3}
  \caption{Comparative summary statistics of the U.S. interconnection queue}
  \label{tab:table_1}
  \centering
  \begin{threeparttable}
    \begin{tabular}{lcccc}
      \toprule
      \textbf{Metric} & \textbf{Value} & \textbf{Basis} &
      \textbf{Cohort} & \textbf{Source} \\
      \midrule

      Active projects (EOY 2025)
      & $\sim$8,513
      & \textit{Projects}
      & Active queue
      & Rand et al.~(2026) \cite{rand2026} \\

      Active generation capacity
      & $\sim$1,312 GW
      & \textit{Capacity}
      & Active queue
      & Rand et al.~(2026) \cite{rand2026} \\

      Active storage capacity
      & $\sim$749 GW
      & \textit{Capacity}
      & Active queue
      & Rand et al.~(2026) \cite{rand2026} \\

      Completion rate
      & 13\%\tnote{a}
      & \textit{Capacity}
      & IR 2000--2020
      & Rand et al.~(2026) \cite{rand2026} \\

      Completion rate
      & $\sim$19\%\tnote{b}
      & \textit{Project count}
      & IR 2000--2020
      & Rand et al.~(2026) \cite{rand2026} \\

      Withdrawal rate
      & 75.7\%\tnote{a}
      & \textit{Capacity}
      & IR 2000--2020
      & Rand et al.~(2026) \cite{rand2026} \\

      Still active
      & 10.0\%\tnote{a}
      & \textit{Capacity}
      & IR 2000--2020
      & Rand et al.~(2026) \cite{rand2026} \\

      Median IR-to-COD (recent)
      & $>5$ years\tnote{c}
      & --
      & COD 2024--2025
      & Rand et al.~(2026) \cite{rand2026} \\

      \bottomrule
    \end{tabular}

    \begin{tablenotes}[flushleft]\footnotesize
      \item[a] Capacity shares for \(2000\text{--}2020\) cohort are \(13.0\%\) operational, \(75.7\%\) withdrawn, \(10.0\%\) active, and
      \(1.3\%\) suspended.

      \item[b] Capacity-weighted completion rates (\(13\%\)) differ from
      project-count completion rates (\(\sim19\%\)).

      \item[c] IR denotes interconnection request; COD denotes commercial
      operation date.

      \item[d] On a capacity-weighted basis, larger projects (gas \(17\%\) and wind \(15\%\)) have higher completion rates than smaller ones (solar \(10\%\) and battery storage \(8\%\)). \cite{rand2026,gorman2025grid}.
    \end{tablenotes}
  \end{threeparttable}
\end{table*}

\subsection{Structural Properties Beyond Processing Delay}
\label{subsec:structural_properties_beyond_processing_delay}
\IEEEPARstart{I}{}nterconnection queues have experienced delays due to study delays, staffing constraints combined with high interconnection application volumes, and legacy serial processes that motivate the framing of the queue congestion as a processing speed problem. FERC Order No. 2023 mandates a transition from legacy serial, first-come-first-served interconnection studies to cluster-based, first-ready-first-served \cite{ferc2023}. While existing reforms address speed inefficiencies, the present paper argues that their proposed framing is incomplete and fails to account for the queue vulnerability to cascading withdrawal failure. When a project withdraws from the interconnection queue, its share of upgrade costs is reallocated to the remaining projects, which may subsequently leave, causing further cost reallocation. The queue does not have a specific mechanism in place to interrupt this feedback once initiated. 

The queue couples projects electrically in a shared transmission infrastructure, and thus, creates a tightly interconnected system. A waiting project withdrawal can therefore induce a cascade of restudy obligations and cost redistribution. Once initiated, through the withdrawal-restudy coupling described in Section~\ref{subsec:cascading_failure_mechanisms}, these dynamics can propagate and amplify in the queue system analogous to the non-linear cascading dynamics observed in interconnected power systems \cite{dobson2007complex, carreras2002critical}, setting off further withdrawals that extend across local administrative boundaries and exceed the predictive capacity of entity-level models \cite{perrow1984}. Grid operators have remarked on these cascading dynamics in their regulatory filings. In its 2021 queue reform filing, Midcontinent Independent System Operator (MISO) stated: ''Because generator and transmission assumptions are interrelated, whenever a higher-queued project withdraws from the queue, that withdrawal changes the underlying assumptions in the interconnection studies conducted for lower-queued projects. The resulting restudies, which often have a cascading effect, impair [the] ability to administer the queue in a timely fashion and significantly delay the execution of GIAs by Interconnection Customers.'' \cite{miso2022}. Similarly, Southwest Power Pool (SPP) attributed multi-year delays in the interconnection study process to ''customers' habit of withdrawing projects from the queue late in the game'', noting that ''such withdrawals can shift the responsibility for upgrades to lower queued customers, requiring cascading restudies'' \cite{spp2021}.

Cascading failures in financial and power-grid networks provide parallels to the propagation mechanisms identified in the U.S. interconnection queue system (see the formal mapping in Section~\ref{subsec:formal_mapping_to_cascading_failure_models}, Table~\ref{tab:table_5}). In each case, the failure of one node redistributes its load to connected nodes, some of which then fail with the additional load exceeding their capacity, generating further redistribution. In financial systems, the failure of a single counterparty can initiate margin calls and forced liquidations, as well as contagion that spreads through the network along channels of exposure, as direct obligations between counterparties that transmit losses when one party defaults \cite{buldyrev2010catastrophic, haldane2011systemic}. Power systems themselves show that failure of a single transmission line can trigger protective relay actions and load redistribution, leading to cascading outages that extend far beyond the initial disturbance \cite{dobson2007complex, carreras2002critical}. In the national queue, withdrawal of a single project waiting to access the grid changes the network-flow assumptions underlying the studies of all downstream interdependent projects. Altered assumptions may then render previously viable projects uneconomic through cost reallocation, driving additional withdrawals and restudies in a cost-withdrawal feedback cycle. 

The queue system operates as a tightly coupled interdependent network with cascading failure modes structurally similar to financial and power grid systems. Unlike these systems, it lacks a formal model of how failures propagate in the system, a stress-testing framework, and relevant resilience metrics. 

\subsection{Study Contributions}
\label{subsec:study_contributions}
\IEEEPARstart{F}{}irst, we introduce a framework for analyzing structural queue fragility that is defined as the vulnerability of a tightly coupled interconnection queue system to cascading withdrawal failure. This framework employs resilience engineering principles, in which system-level failures require system-level models \cite{hollnagel2006resilience}. The computational model introduced in this paper adapts two cascade formalisms of the threshold cascade model by Watts (2002) \cite{watts2002simple} and the interdependent-network percolation model by Buldyrev \etal ~(2010) \cite{buldyrev2010catastrophic}. Cost reallocation is modeled using an Eisenberg--Noe-type clearing mechanism \cite{eisenberg2001systemic}. Second, we formalize four queue-specific properties that generate the empirically observed clustering and amplification dynamics that we refer as withdrawal-restudy coupling, cost concentration risk, temporal clustering of failures, and regulatory feedback loops. Third, we map cascading failure models from financial and interdependent infrastructure systems onto the queue system (Table~\ref{tab:table_5}), providing a formal basis for applying existing theory to queue withdrawal dynamics. Fourth, initial empirical evidence is presented for cascading dynamics, derived from temporal clustering analysis and burst identification applied to project-level withdrawal records across all seven ISO/RTOs (Section~\ref{sec:preliminary_empirical_analysis}). Finally, we develop a stylized computational model to analyze the cascade multiplier effect, which shows that amplification arises from the structural properties identified in the paper. The computational model is then used to simulate and test the performance of the targeted circuit-breaker interventions (Section~\ref{sec:computational_contagion_model_of_queue_fragility}).

\subsection{Security Considerations}
\label{subsec:security_considerations}
\IEEEPARstart{T}{}he fragility of the interconnection process now poses national security risks, extending beyond its direct effects on energy deployment. Hyperscale AI data centers, requiring hundreds of megawatts to multiple gigawatts of reliable power supply in demand, have introduced a load type the interconnection process was not designed to accommodate. Regardless of strategic destination, every such facility must complete the same interconnection study process that every other applicant needs to navigate. In such a complex environment, multi-gigawatt power campus plans have been announced by major technology companies, with individual facilities requiring sustained supply in the \(100\) MW\(\text{--}1\) GW range. Their timelines are affected by the cascading dynamics described in Section~\ref{sec:the_interconnection_queue_as_a_complex_adaptive_system}. Federal policy has already treated AI infrastructure as a strategically significant asset \cite{whitehouse2024a, whitehouse2025a, whitehouse2025b} that highlights the implications of queue fragility extending beyond the energy sector. 

The energy infrastructure has become foundational for both economic competitiveness and national security. Facilities that host defense computation and civilian AI systems pass through the same fragile queue system described here, and they require reliable, newly constructed generation capacity. Failure to reliably deliver the new generation capacity, therefore, constitutes a macro-structural vulnerability. Following incidents between 2024 and 2025, when more than a gigawatt of data-center load disconnected from the grid within seconds, North American Electric Reliability Corporation (NERC) escalated a Level 2 Alert on large loads (September 2025) to a Level 3 Essential Action Alert (May 2026), as the only the third Level 3 Alert in its 58-year history. NERC subsequently initiated ''Project 2026-02'' and introduced a mandatory registration category for large computational loads \cite{nerc2026}. Two regulatory institutions of FERC on access and process and NERC on operations point toward a common concern from different directions. The interconnection system can generate system-level failure modes that are not well captured by regulations focused on individual entities. Neither regulatory approach, however, is yet informed by a formal model of how such failures propagate through the broader queue system. Capacity is leaving the system faster than new capacity can arrive, with projected generator retirements totaling \(115\) GW between 2025 and 2034, and resource additions falling short of industry projections for two consecutive cycles \cite{nerc2024}. Capacity auction costs of Pennsylvania-New Jersey-Maryland Interconnection (PJM) raised from \$\(2.2\) billion to \$\(14.7\) billion between the 2024/2025 and 2025/2026 delivery years \cite{pjm2024}, driven by changes in resource accreditation methodology, generator retirements, and growing electricity demand. A national congested queue that limits new resource entry is now an additional contributing factor. Johnston, Liu and Yang (2023) \cite{johnston2023empirical} find high interconnection costs as a primary driver of generator withdrawal decisions. These costs are ultimately borne by nearly 67 million people in the service territory of PJM across 13 states and the District of Columbia \cite{pjm2024}.




\section{Limitations of Current Approaches}
\label{sec:limitations_of_current_approaches}

\subsection{FERC Order 2023 and Cluster Study Reforms}
\label{subsec:ferc_order_2023_and_cluster_study_refroms}
\IEEEPARstart{F}{}ERC Order 2023 restructures serial, first-come-first-served interconnection request processing into cluster-based, first-ready-first-served processing. Its primary provisions require cluster study timelines (a 45-day customer request window, a 60-day customer engagement window and a 150-day cluster study period). Order 2023 enforces stricter financial readiness requirements of increasing deposits and penalties for projects that withdraw from the queue. It also introduces penalties for transmission providers that miss study deadlines and provisions for affected system planning \cite{ferc2023, ferc2024}.

By addressing the serial study process, the current reforms target a specific structural failure mode. A queue entry does not require site control or secured financing, allowing developers to speculatively hold network capacity positions, a form of options behavior in which developers monetize queue priority and displace viable projects that have secured the financing and other requirements, while avoiding commitment costs. When these speculative entries eventually withdraw, they initiate iterative restudies propagating through the entire system and increasing timeline uncertainty (Section~\ref{subsec:structural_properties_beyond_processing_delay}). The speculative positions also increase the share of non-viable projects in the queue composition and distort its alignment with developable generation capacity. By requiring a queue entry to establish verifiable financial readiness, the cluster approach aims to reduce speculative project submissions and subsequent withdrawals and restudies that they trigger.

The current reforms failed to target network-level interdependence, and focus on the process-level symptoms. Portfolio decisions remain structurally disconnected from network feasibility assessments. As a result, simply accelerating cluster study throughput does not prevent failure propagation, if their interdependency is ignored. The 2025 Expedited Resource Addition Study (ERAS) proposal by MISO explicitly acknowledged that ''minimizing the restudy risk that is inherent to multi-phase cluster studies'' required moving to serial studies for expedited projects specifically to avoid ''the risk of cascading restudies'' \cite{miso2025a}. That MISO reverted to serial studies for priority projects signals a limit of cluster approaches.

Regulatory developments after Order 2023 have accelerated. In October 2025, the Secretary of Energy directed FERC to consider regulation on large-load interconnection. In June 2026, FERC then issued show cause orders under Section 206 of the Federal Power Act to all six jurisdictional ISO/RTOs with two deadlines: a 30-day informational report addressing regional resource adequacy for large loads, and a 60-day filing either defending existing tariff provisions or proposing reforms ~\cite{ferc2026a}. Five operators then requested the temporary suspension of the 60-day deadline; FERC granted the requests, capped them at 90 days. The open dockets therefore broaden the resilience-engineering gap discussed in the present paper, and create an administrative mechanism, through which the interventions proposed in Section~\ref{sec:policy_and_national_security_dimensions} could be considered. 

Using PJM data, Johnston, Liu and Yang (2023) \cite{johnston2023empirical} analyze how queue congestion and interconnection costs interact to drive project withdrawal decisions. Google/Alphabet X’s Tapestry initiative, in partnership with PJM, develops grid visibility tools. Other endeavors address different parts of the interconnection problem. Federal initiatives, particularly Interconnection Innovation e-Xchange (i2X) program of the Department of Energy, target application-processing efficiency through automation and data infrastructure improvement. These initiatives collectively contribute to grid modernization; however, they address a failure mode different from that of systemic resilience. A queue may become more transparent and more efficiently administered, while remaining vulnerable to cascading failure. We focus on the lack of a resilience framework for managing the interdependencies in the queue system.

\subsection{Resilience Engineering Gap}
\label{subsec:resilience_engineering_gap}
\IEEEPARstart{A}{}s of mid-2026, a review of existing reforms reveals that there is no formal approach treating the U.S. interconnection queue as a complex system that needs engineered resilience. The resilience engineering framework \cite{hollnagel2006resilience} and systems safety \cite{leveson2011engineering}, addressing complex socio-technical systems, have not yet been applied to the queue system. Although current reforms address speculative entry, slow studies, and insufficient staffing, they do not engage with the cascading failure vulnerability in the queue process. Designed for a grid with relatively few large conventional generators arriving at a predictable pace, the queue served as a simple administrative process that did not require this engineering approach. Today, however, the queue processes thousands of simultaneous requests from intermittent resources, storage systems, and hybrid facilities. The environment has changed so much that the queue now behaves like a complex system susceptible to systemic failure.

Gorman \etal ~(2025) \cite{gorman2025grid} points to assessed interconnection costs for withdrawn projects average \$\(373/\)kW, more than five times the \$\(73/\)kW average for completed projects. Johnston, Liu and Yang (2023) \cite{johnston2023empirical} model the withdrawal as an ''optimal stopping problem'', in which developers form expectations about study arrival and interconnection costs within a non-stationary queuing equilibrium framework. This approach allows for a detailed analysis of queue withdrawal behavior at the individual developer level. Their framework is complementary to, rather than competing with, the framework proposed in the present paper. The former focuses on the optimization problem faced by the individual developers, and the latter focuses on characterizing network dynamics; specifically, how developers' exits affect the conditions faced by other electrically interdependent projects. Their equilibrium exit rule could, in principle, provide a microfoundation for the viability threshold used in Section~\ref{sec:computational_contagion_model_of_queue_fragility}. Both approaches, however, ultimately treat aggregate outcomes differently, where agent-optimization models explain who withdraws, and the contagion framework explains why withdrawal cluster, amplify, and propagate. We explore their findings under engineering queue resilience to examine whether aggregate patterns they document, including high attrition, escalating costs and temporal clustering, are emergent behaviors of the queue system whose architecture makes cascading failure the expected outcome. We propose a shift from the process-oriented to system-level perspective, which makes resilience engineering of the queue possible.



\section{The Interconnection Queue as a Complex Adaptive System}
\label{sec:the_interconnection_queue_as_a_complex_adaptive_system}

\subsection{Theoretical Foundations}
\label{subsec:theoretical_foundations}
\IEEEPARstart{W}{}e apply the Complex Adaptive System (CAS) framework to the interconnection queue aimed to examine the system-level cascade dynamics. Holling (1973) \cite{holling1973} defines system resilience as disturbance-absorptive capacity while maintaining function, extended by Folke (2006) \cite{folke2006resilience} to socio-ecological systems. For this paper, complex network theory provides analytical tools for characterizing the topology of interdependent systems \cite{strogatz2001exploring, newman2003structure}. A complex adaptive system is characterized by: (1) heterogeneous agents having different objectives and decision rules; (2) nonlinear interactions between agents; (3) adaptation, whereby agents modify their behavior in response to system-level outcomes; (4) emergence, whereby system-level patterns arise from agent interactions and are not predictable from individual agent behavior; and (5) path dependence, whereby the system’s trajectory is influenced by its history \cite{holland1995, miller2009complex}. In the interconnection queue, network-upgrade cost allocations can depend on the projects studied previously. Each study is based on assumptions established in preceding studies; therefore, the withdrawal or modification of an earlier project may alter the upgrade requirements and cost allocations for subsequent project.

Under this framework, the interconnection queue satisfies all five complex adaptive system criteria. Its agents include renewable energy developers (solar, wind, storage, hybrid), conventional generators (natural gas, nuclear), large-load customers (data centers, industrial facilities), transmission providers, and regulators. Agent heterogeneity spans cost-based constraints, risk tolerances, regulatory obligations and objectives. Their interactions are nonlinear, because a single project withdrawal may initiate disproportionate cost reallocations and restudy cascades depending on the position, technical characteristics, and size of a withdrawing project. In response to observed dynamics, developers modify their queue entry strategies, site selection, and financial structuring, and thus, agents adapt. At approximately \(76\%\), the capacity withdrawal rate \cite{rand2026} is not sufficient, on its own, to explain the observed agent behavior, and suggests an emergent structural property in the queue system.

\subsection{Cascading Failure Mechanisms}
\label{subsec:cascading_failure_mechanisms}
\begin{figure*}[!t]
  \centering
  \includegraphics[width=6.5in]{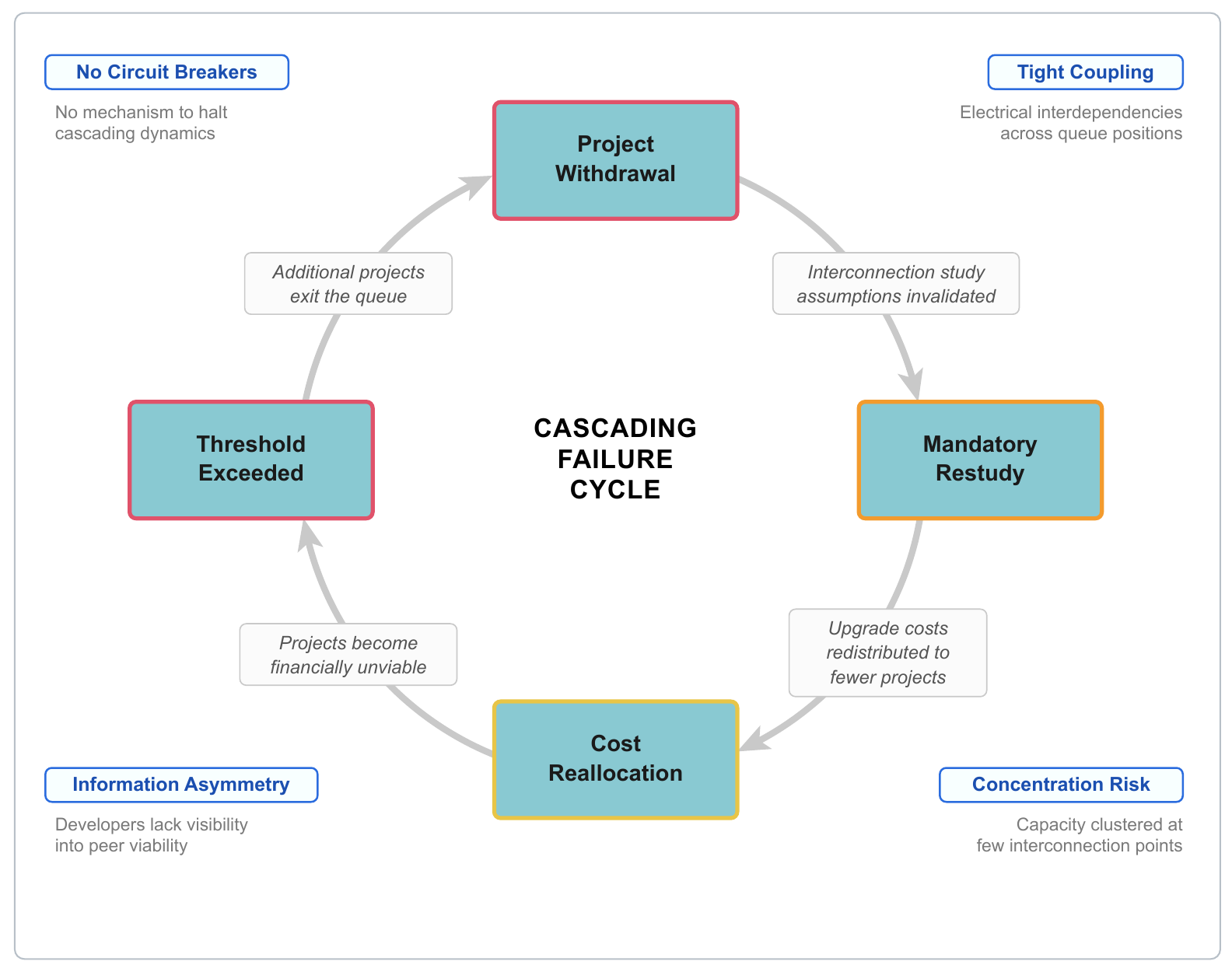}
  \caption{Failure propagation mechanism of the U.S. interconnection queue}
  \label{fig:figure_1}
\end{figure*}

\IEEEPARstart{F}{}igure~\ref{fig:figure_1} plots four failure propagation mechanisms and their associated feedback relationships. The mechanisms depict coupling between project withdrawal and restudy, amplification effect resulting from cost reallocation, temporal clustering, and regulatory feedback loops.

Once a project withdrawn, the network-flow assumptions underpinning interdependent studies no longer hold. This prompts the transmission providers to conduct restudies and reassess the remaining projects against the altered assumptions. Electrical interdependence determines the scope of restudy, whether dozens or hundreds of projects across multiple study clusters and voltage levels.

A project withdrawal can invalidate the cost-sharing assumptions of the original study. Restudies frequently reveal that network upgrade costs previously shared among many projects now redistribute among fewer remaining projects (consistent with concentration risk in Section~\ref{subsec:built_in_fragility_characteristics}). That reallocation can increase costs sharply and, in some cases, turns previously viable projects into uneconomic projects. In transmission-constrained regions, network upgrade costs have risen substantially. The cost differential between withdrawn projects (\$\(373/\)kW) and completed projects (\$\(73/\)kW) on average by a factor of roughly five \cite{gorman2025grid} corresponds to the cost-concentration mechanism described in Section~\ref{subsec:built_in_fragility_characteristics}. When cost redistribution increases the allocated costs of a project beyond the financial viability threshold of a project, the project withdraws. As subsequent restudies follow and costs shift again, the next withdrawal is thereby established.

Without a uniform distribution over time, withdrawals tend to cluster around specific dates, for example, study milestone dates, restudy completion announcements, and financial commitment deadlines, when developers receive new information about costs. This concentration of withdrawals at decision points creates shocks that exceed the absorption capacity of the affected projects in the queue, amplifying cascade propagation (see Section~\ref{sec:computational_contagion_model_of_queue_fragility}). Each withdrawal alters the existing assumptions, which in turn triggers another round of restudies (see the formal cascade model in Section~\ref{sec:computational_contagion_model_of_queue_fragility}).

Higher deposits that are designed to screen out non-committed applicants also raise the penalty for remaining in a queue amid a spreading reallocation of costs, potentially prompting earlier withdrawals. Expedited processes exist to fast-track priority projects. However, in this situation, they may redirect analytical resources away from standard queue processing, and in turn, extend timelines for non-priority projects and increase their attrition risk. These second-order effects have been largely absent from regulatory impact assessments.

\subsection{Built-in Fragility Characteristics}
\label{subsec:built_in_fragility_characteristics}
\IEEEPARstart{W}{}e identify the following structural properties of the queue that make it vulnerable to a systemic failure: tight coupling, concentration risk, information asymmetry, and the absence of circuit-breakers.

\textbf{\textit{Tight Coupling.}} Regardless of their position in the queue or chronological order, projects are coupled by electrical proximity associated with their shared transmission infrastructure, common network upgrade requirements, and interdependent electrical studies. Cost and restudy disturbances consequently travel through the system along channels invisible in the administrative structure.

Perrow (1984) \cite{perrow1984} identifies tight-coupling, interactively complex systems prone to ‘normal accidents’: inevitable failures due to the architecture of the system itself, not its component malfunction. The interconnection queue process falls into this category. Shared transmission infrastructure links, once a project withdraws, immediately change the cost assumptions for electrically adjacent projects. The effects travel along the interconnection queue network beyond the administrative boundaries, making interactions in the queue complex. A withdrawal in one cluster can therefore result in restudies in another through shared transmission constraints. Perrow (1984) \cite{perrow1984} implies that incremental improvements in the process will not eliminate the failure mode and the coupling architecture itself needs to be redesigned. The interconnection relationship does not simply terminate once contractual commitment has been fulfilled. Among projects from 2000 to 2022, \(41\%\) of capacity with executed interconnection agreements subsequently withdrew (Section~\ref{subsec:interpretation}). This indicates that even after completing initial studies and signing contract, projects may still face cost reallocation and restudy exposure.

\textbf{\textit{Concentration Risk.}} From both geographic and electrical standpoints, the queue capacity is concentrated at a relatively small number of points of interconnection and transmission corridors. When larger projects withdraw from a highly concentrated interconnected node, the remaining smaller projects often face disproportionately higher cost reallocation associated with the interconnection queue. The counterparty concentration characteristic of financial networks is a parallel to this form of concentration risk, where the failure or withdrawal of a highly connected node can generate cascading effects throughout the entire system \cite{crucitti2004model}.

\textbf{\textit{Information Asymmetry}}. Because cluster study processes are not transparent, developers cannot determine the financial viability and withdrawal probability of peer projects in the study cluster, and entry and exit decisions are based on incomplete system-state information. 
Financially marginal developers can enter the queue at relatively low costs and leave without comparable penalty, while developers with viable projects are most likely to remain in the queue long enough to absorb redistribution of shared upgrade costs when less viable projects withdraw. They bear the risk of repricing more. Over time, this mechanism may screen out projects with higher likelihood of withdrawal, and subsequently, concentrate the reallocation risk among the developers with remaining projects in the queue.

\textbf{\textit{Absence of Circuit-breakers}}. No mechanism currently exists within the U.S. interconnection queue to suspend, or otherwise interrupt, cascading dynamics. Financial markets have implemented circuit-breakers to stop trading during cascading sell-offs. However, when a withdrawal cascade begins in the queue system, it continues until the system reaches a new equilibrium due to the lack of similar interventions and risks the withdrawal of a substantial fraction of the affected projects before equilibrium is restored. Gai \& Kapadia (2010) \cite{gai2010contagion} describe such systems as ''robust yet fragile'' that are stable under routine conditions but vulnerable to rare cascading events.

\subsection{System Analogies}
\label{subsec:system_analogies}
\IEEEPARstart{R}{}esearch on network contagion effects during the 2008 financial crisis suggests that highly interconnected networks, particularly those characterized by a high counterparty concentration, may amplify cascading effects, leading to consequences that extend far beyond the initial shock \cite{haldane2011systemic}. Post-crisis regulatory responses have primarily included conducting stress tests, imposing capital adequacy requirements, and systemic risk monitoring interventions \cite{haldane2011systemic, acemoglu2015systemic, eisenberg2001systemic}. These analytical tools can be adapted to the interconnection queue. We borrow from the established financial contagion models and adapt them to the interconnection queue system.

Buldyrev \etal ~(2010) \cite{buldyrev2010catastrophic} and Gao \etal ~(2012) \cite{gao2012networks} show that strong interdependencies result in phase transitions between localized failure and system-wide collapse. Carreras \etal ~(2002) \cite{carreras2002critical}, Dobson \etal ~(2007) \cite{dobson2007complex} and Soltan, Mazauric, \& Zussman (2015) \cite{soltan2015analysis} add that analyses of power grid blackouts exhibit self-organized criticality with cascade size distributions following power-law scaling. The queue may cross the same threshold by shifting from scattered withdrawals to systemic cascades.

Thousands of simultaneous operations are coordinated in modern air traffic control through continuous system-wide optimization, accounting for interdependencies between aircraft, runways, and weather patterns. By contrast, a queue system processes a comparable number of concurrent requests without a system-level resilience design.

\subsection{Formal Mapping to Cascading Failure Models}
\label{subsec:formal_mapping_to_cascading_failure_models}
\IEEEPARstart{T}{}he framework integrates the principles of interdependent-network percolation \cite{buldyrev2010catastrophic, gao2012networks}, threshold cascade models \cite{watts2002simple, motter2002cascade}, and financial contagion models \cite{acemoglu2015systemic, eisenberg2001systemic, gai2010contagion}. Table~\ref{tab:table_5} provides a formal mapping between the key elements of these cascading-failure models and the interconnection queue system. The inter-project cost-sharing dependencies function as interdependent network links reported by Buldyrev \etal ~(2010) \cite{buldyrev2010catastrophic}. In this case, the developer's decision to withdraw is interpreted as a threshold response established on a Watts-type threshold rule. The subsequent reallocation of upgrade costs follows an Eisenberg--Noe-type clearing mechanism to the extent that the ''payment'' is the upgrade cost obligation that must be absorbed by surviving projects.

The mappings are approximate. Unlike bilateral dependencies between two distinct infrastructure networks described by Buldyrev, the interconnection queue contains multilateral interdependencies. A project withdrawal impacts all electrically interconnected projects simultaneously within the network rather than only pairs of dependent nodes. In the standard Watts’ formulation, thresholds are static, and developer viability thresholds shift with market conditions, policy changes, and project timelines. All nodes experience Eisenberg--Noe clearing process simultaneously, while queue restudies proceed sequentially over months. Projects receive revised cost allocations at different times.

Phase-transition behavior is the most directly relevant theoretical finding from financial contagion literature. Acemoglu \etal ~(2015) \cite{acemoglu2015systemic} state that denser interconnections enhance stability under small shocks but amplify contagion for large ones. Cost allocations that are shared across more projects in the queue distribute the cost of individual withdrawals (stabilizing for small shocks) but also multiply channels through which cascades spread (destabilizing for large shocks). Our numerical model in Section~\ref{sec:computational_contagion_model_of_queue_fragility} tests whether this phase transition appears in a stylized Erd\H{o}s--R\'enyi network. The theoretical prediction by Acemoglu \etal ~(2015) \cite{acemoglu2015systemic} for this phase-transition dynamic points to a parallel in the queue’s behavior at high connectivity. Battiston \etal ~(2012) \cite{battiston2012liaisons} find that the same diversification reducing individual risk can simultaneously increase systemic risk through contagion channels, which has a direct analogue to the cluster study approach of FERC Order 2023. Although studying projects together increases efficiency and distributes costs, it also increases the coupling through which cascading dynamics operate.

\begin{table*}[!t]
  \renewcommand{\arraystretch}{1.3}
  \caption{Formal mapping of cascading failure models onto the interconnection queue process}
  \label{tab:table_5}
  \centering
  \begin{threeparttable}
    \begin{tabular}{L{2.2cm} L{3.0cm} L{3.0cm} L{3.0cm} L{3.4cm}}
      \toprule
      \textbf{Component} &
      \textbf{Buldyrev \etal ~(2010) \cite{buldyrev2010catastrophic}} &
      \textbf{Watts (2002) \cite{watts2002simple}} &
      \textbf{Acemoglu \etal ~(2015) \cite{acemoglu2015systemic}} &
      \textbf{Queue Mapping} \\
      \midrule

      Node
      & Infrastructure component
      & Agent with binary state
      & Financial institution
      & Queued project \\

      Failure trigger
      & Loss of dependent node
      & Neighbor fraction $>$ threshold
      & Losses $>$ equity buffer
      & Cost reallocation $>$ viability threshold \\

      Edge
      & Physical dependency
      & Social/information link
      & Interbank liability
      & Shared upgrade cost allocation \\

      Phase transition
      & Percolation threshold
      & Cascade window
      & Small $\rightarrow$ large shock transition
      & Critical connectivity for cluster collapse \\

      Intervention
      & Reduce dependencies
      & Increase threshold diversity
      & Capital adequacy buffers
      & Circuit-breaker: cap cost redistribution \\

      \bottomrule
    \end{tabular}
  \end{threeparttable}
\end{table*}

We develop a framework based on a two-part hypothesis. Part (i) proposes that, under current conditions, the queue system behaves similarly to a contagion system, exhibiting a moderate amplification effect. Withdrawal shocks propagate through cost-sharing interdependencies. Net amplification becomes greater than unity but well below the threshold of runaway systemic collapse. Part (ii) proposes that the transition to a state of systemic collapse depends on two conditions being satisfied simultaneously: high interdependence of cost-sharing and exposure to the risk of non-dilutive cost per project. If either condition is not met, the system collapse does not occur. Sections~\ref{sec:preliminary_empirical_analysis} and~\ref{sec:computational_contagion_model_of_queue_fragility} test these propositions using empirical and computation methods. 



\section{Preliminary Empirical Analysis}
\label{sec:preliminary_empirical_analysis}

\subsection{Data Source and Methods}
\label{subsec:date_source_and_methods}
\IEEEPARstart{F}{}or the empirical analysis, we use the LBNL Queued Up 2026 Edition with \(38\text{,}201\) project-level records from more than 50 transmission operators \cite{rand2026}. We analyze seven ISO/RTO regions of Pennsylvania-New Jersey-Maryland (PJM), the Southwest Power Pool (SPP), Midcontinent Independent System Operator (MISO), the Electric Reliability Council of Texas (ERCOT), ISO New England (ISO-NE), the California Independent System Operator (CAISO), and the New York Independent System Operator (NYISO). The empirical analysis accounts for \(25\text{,}770\) projects in total, including \(16\text{,}206\) withdrawals. 

The data on withdrawal dates are only reported for \(13\text{,}984\) out of 16,206 withdrawn projects (\(86.3\%\)). Data coverage also varies across the seven regions: \(98.4\%\) for PJM, \(48.7\%\) for ERCOT, \(98.8\%\) for ISO-NE, \(98.2\%\) for CAISO, \(72.8\%\) for SPP, \(48.7\%\) for NYISO, and \(93.8\%\) for MISO. The relatively low coverage rates for ERCOT and NYISO raise concern that withdrawal dates were recorded selectively. We retain all seven regions in the temporal clustering analysis and explicitly identify the estimated dispersion indices for ERCOT and NYISO as lower bounds in Table~\ref{tab:table_2}.

We use the dispersion index \(D\) to evaluate temporal clustering, which is defined as the variance-to-mean ratio of monthly withdrawal counts for each ISO/RTO from 2010 to 2025. Under the Poisson null hypothesis, assuming that random withdrawal events occur independently over time, the expected value of dispersion statistic \(D\) is 1. A value greater than 1 is interpreted as overdispersion, which is consistent with temporal clustering \(D > 1\). We use a chi-squared goodness-of-fit test (\(\alpha = 0.001\); \(df = 191\)) to test data deviation from that benchmark. Table~\ref{tab:table_2} lists the dispersion indices computed relative to the mean over the complete study period. To adjust for the long-term growth in the total number of projects in the queue over the study period, we also recompute dispersion indices relative to within-year mean for each individual calendar year (Table~\ref{tab:table_2} note).

We then use withdrawn capacity to identify withdrawal bursts. For each ISO/RTO, a month is classified as experiencing a withdrawal burst if its withdrawn capacity exceeds the regional mean by \(\mu + 2\sigma\), where \(\mu\) and \(\sigma\) denote respectively the regional mean and standard deviation of monthly withdrawn capacity (GW) over the study period. In addition, we examine the distribution of withdrawal severity using the Complementary Cumulative Distribution Function (CCDF) of monthly withdrawal ratios. For each ISO/RTO, this ratio is defined as the withdrawn capacity in a given month divided by mean monthly withdrawn capacity of that ISO/RTO over the study period. We then pool the normalized values from the seven regions (\(n = 1{,}344\) region-months). The dispersion index is computed using withdrawal counts in order to characterize the temporal occurrence of withdrawal events, and identification of withdrawal burst and the CCDF analysis are based on withdrawn capacity aimed to characterize event severity. We plot the CCDF on logarithmic probability axes and compare it with a reference exponential distribution in order to evaluate heavy-tailed behavior. 

We then estimate the conditional probability of co-withdrawal and compare it with a permutation null hypothesis. For each withdrawn project \(i\), we define its cohort peer set as, 

\[
\begin{aligned}
P(i)=\{\, j \neq i \mid\;& r(j)=r(i),\\
& \tau(j)=\tau(i),\\
& \lvert e(j)-e(i)\rvert \leq 6 \,\}.
\end{aligned}
\]

where \(r\) denotes the ISO/RTO that a record belongs to, \(\tau\) is the technology category recorded in the LBNL data, and \(e\) indicates the interconnection-request (entry) month. We consider the project membership of a cohort if they belong to the same region and technology category, and if the dates on which they entered the queue fall within \(\pm 6\) calendar months of one another. We evaluate each project within its own peer set. For project \(i\), the co-withdrawal indicator is defined as, 

\[
C(i)=\mathbf{1}\!\left[
\exists\, j \in P(i) :
\lvert w(j)-w(i)\rvert \leq 12
\right].
\]

where \(w\) denotes the month when a withdrawal occurs. \(C(i) = 1\) if at least one other project in the cohort withdraws within \(\pm 12\) calendar months before or after the withdrawal date of the project \(i\). The fraction of withdrawn projects satisfying the aforementioned condition is determined by averaging \(C(i)\) across all withdrawn projects in the region. All entry- and withdrawal-time differences are measured in calendar months.

The LBNL Queued Up 2026 Edition data contains 53 withdrawal dates after 31 December 2025, which we consider outside our analysis window and are therefore excluded. 665 records lack an entry date, six record a withdrawal date preceding the entry date, 257 withdrawn projects carry a commercial operation date, and 14 operational projects carry a withdrawal date. We excluded any records with internally inconsistent date combinations from the date-based analyses. 

The cohort peer set is an observable proxy for identifying projects that may have been exposed to common cost-reallocation conditions. It must not be interpreted as representation of electrical connections or cost-sharing adjacency depicted by the network edges of the model in Section~\ref{sec:computational_contagion_model_of_queue_fragility}. Direct reconstruction of those relationships requires the records of cost allocation at the project level. 

Two permutation nulls are used. An unrestricted null reassigns withdrawal dates at random among withdrawn projects within each ISO/RTO. We report the unrestricted approach as a baseline for comparison and is not recommended as standalone. Because entry and withdrawal months are strongly correlated in the queue across all seven regions (Pearson \(r = 0.72\text{--}0.96\)), unrestricted reassignment produces permuted datasets. In this case, roughly \(35\%\) of projects withdraw before they enter, and it therefore eliminates entry-to-exit ordering rather than only cohort-specific timing. To correct this, we employ two constrained nulls. The first stratifies the permutation by entry year, reassigning withdrawal dates only among projects that entered the queue in the same calendar year and preserving the entry-exit relationship. The second permutes the technology label and holds both dates fixed. This preserves all temporal structure and tests only whether co-withdrawal is concentrated within technology categories. All three permutation null models are simulated across 1,000 permutations. By retaining the aggregate temporal clustering present within each region, we determine co-withdrawal associated with cohorts or technology categories; that is, clustering beyond the aggregate level captured by the dispersion index. We thereby identify correlations between the withdrawal timing and cohort dimensions. 

\subsection{Empirical Analysis Results}
\label{subsec:empirical_analysis_results}
\IEEEPARstart{T}{}able~\ref{tab:table_2} summarizes the dispersion indices and chi-squared statistics for the seven ISO/RTOs. All regions show statistically significant overdispersion indices (\(\chi^2(191) = 947\text{--}19{,}559\), all \(p < 0.001\)), ranging from 4.96 for NYISO to 102.40 for CAISO. The dispersion indices vary substantially across the regions. CAISO has the highest index (\(D = 102.40\)), largely attributable to the December 2024 ''cluster-deadline'' event discussed below. ISO-NE (\(68.11\)), MISO (\(67.48\)), and PJM (\(62.85\)) also show high dispersion indices, while SPP has a comparatively lower index \(34.28\). ERCOT (\(6.83\)) and NYISO (\(4.96\)) have the lowest indices. Because coverage of withdrawal dates for these two regions is below \(50\%\) in the LBNL Queued Up 2026 Edition dataset, their values are regarded as lower bounds. For all seven ISO/RTOs, the Poisson null hypothesis that withdrawals occur randomly and are uniformly distributed over time is rejected (\(\chi^2(191) \geq 947\), \(p < 0.001\)). Given the magnitude of overdispersion and its variation observed during restudy and cost-reallocation cycles, the observed temporal clustering is unlikely to arise from sampling variation alone (Section~\ref{subsec:interpretation}). The chi-squared test establishes overdispersion relative to the Poisson benchmark. Since trend or seasonality alone would produce overdispersion, identification of cohort- and technology-specific clustering primarily relies on the permutation tests reported below and on the burst analysis. 

\begin{table*}[!t]
  \renewcommand{\arraystretch}{1.25}
  \caption{Withdrawal temporal clustering of the interconnection queue by ISO/RTO between 2010 and 2025}
  \label{tab:table_2}
  \centering
  \begin{threeparttable}
    \begin{tabular}{lrrrrrrc}
      \toprule
      \textbf{ISO/RTO} &
      \textbf{\(N\)} &
      \textbf{Monthly Mean} &
      \textbf{Monthly Var.} &
      \textbf{Disp. Index} &
      \textbf{\(\chi^2\) (df = 191)} &
      \textbf{\(p\)-value} &
      \textbf{Sig.} \\
      \midrule

      PJM
      & 4{,}649
      & 24.21
      & 1{,}521.81
      & 62.85
      & 12{,}004
      & \(<0.001\)
      & *** \\

      MISO
      & 2{,}682
      & 13.97
      & 942.68
      & 67.48
      & 12{,}890
      & \(<0.001\)
      & *** \\

      CAISO\tnote{a}
      & 1{,}909
      & 9.94
      & 1{,}018.17
      & 102.40
      & 19{,}559
      & \(<0.001\)
      & *** \\

      ERCOT\tnote{b}
      & 564
      & 2.94
      & 20.07
      & 6.83
      & 1{,}305
      & \(<0.001\)
      & *** \\

      ISO-NE
      & 894
      & 4.66
      & 317.13
      & 68.11
      & 13{,}009
      & \(<0.001\)
      & *** \\

      SPP
      & 1{,}326
      & 6.91
      & 236.72
      & 34.28
      & 6{,}547
      & \(<0.001\)
      & *** \\

      NYISO\tnote{b}
      & 658
      & 3.43
      & 17.00
      & 4.96
      & 947
      & \(<0.001\)
      & *** \\

      \bottomrule
    \end{tabular}

    \begin{tablenotes}[flushleft]\footnotesize
      \item[a] In CAISO, the December 2024 withdrawal burst alone
      accounts for \(89\%\) of the region's monthly variance.

      \item[b] 2026 withdrawal-date coverage is below \(50\%\) (ERCOT and NYISO have \(48.7\%\) coverage).

      \item[c] The dispersion index is defined as
      \(D=\mathrm{Variance}/\mathrm{Mean}\) of monthly withdrawal counts
      from 2010 through 2025. Under the Poisson null hypothesis of
      randomly occurring withdrawals, \(D=1\).

      \item[d] The goodness-of-fit statistic is
      \(\chi^2=(n-1)D\), where \(n=192\) monthly observations and
      \(\mathrm{df}=191\). The critical value is
      \(\chi^2_{0.001,191}=257.1\).
      All reported values satisfy \(p<0.001\).

      \item[e] A small number of months with extreme values
      account for much of the observed dispersion. Overdispersion
      persists when the dispersion index is computed within calendar
      years, thereby controlling for the upward trend in the number of
      queued projects over time: median within-year D ranges from \(1.53\) (ISO-NE) to \(9.69\) (CAISO), and four regions (ERCOT, ISO-NE, SPP, NYISO) contain individual years indistinguishable from the Poisson benchmark (minimum single-year \(D = 0.74\)).

      \item[f] Data: Rand et al.~(2026) \cite{rand2026}.
    \end{tablenotes}
  \end{threeparttable}
\end{table*}

The dispersion indices are sensitive to a small number of extreme months. Removing the single largest month of each region cuts the index sharply where dispersion results from one event (CAISO, \(102.40\) to \(13.87\)) (Table~\ref{tab:table_3}). With three largest months removed in each region, the indices re-range from \(3.76\) (ISO-NE) to \(29.66\) (MISO), all significantly above the Poisson benchmark of unity (\(p < 0.001\)).

In CAISO's December 2024 burst, \(420\) out of \(426\) records carry the date 2 December 2024, and \(303\) belong to a single Cluster 15 identifier block (\(132.1\) GW) for which county information is absent; the largest contribution of ISO-NE is \(233\) records sharing 13 June 2025, totaling only \(2.35\) GW, with queue-entry dates spanning 2000 to 2025. PJM December 2024 burst includes \(407\) records sharing 18 December 2024. Excluding the 303 imputed Cluster 15 identifiers reduces CAISO's index to \(21.08\); excluding the full \(420\)-record Cluster 15 block gives \(13.82\). Excluding the Cluster 15 block and then removing every date on which 100 or more records in a region share a single withdrawal date (six dates, total of \(1\text{,}206\) records) yields indices of \(4.96\) (NYISO) to \(34.28\) (SPP), all far above unity. These records indicate administrative and cluster-level date assignment practices and do not establish same-day withdrawal decisions by independent developers. 
Nevertheless, temporal clustering survives both checks, and is considered a property of the withdrawal series rather than of outlier months or recording practice. 

\begin{table*}[!t]
  \renewcommand{\arraystretch}{1.3}
  \caption{Sensitivity of the dispersion index to extreme months and to date-assignment convention, 2010--2025}
  \label{tab:table_3}
  \centering
  \begin{threeparttable}
    \begin{tabular}{lcccc}
      \toprule
      \textbf{ISO/RTO} &
      \textbf{All Months} &
      \textbf{Drop Largest} &
      \textbf{Drop 3 Largest} &
      \textbf{Excluding Imputed Dates} \\
      \midrule

      CAISO
      & 102.40
      & 13.87
      & 11.20
      & 13.82 \\

      ISO-NE
      & 68.11
      & 8.24
      & 3.76
      & 8.17 \\

      MISO
      & 67.48
      & 56.09
      & 29.66
      & 32.23 \\

      PJM
      & 62.85
      & 18.64
      & 13.54
      & 19.05 \\

      SPP
      & 34.28
      & 24.82
      & 12.49
      & 34.28 \\

      ERCOT\tnote{a}
      & 6.83
      & 6.16
      & 5.51
      & 6.83 \\

      NYISO\tnote{a}
      & 4.96
      & 4.58
      & 4.21
      & 4.96 \\

      \bottomrule
    \end{tabular}

    \begin{tablenotes}[flushleft]\footnotesize
      \item[a] Withdrawal-date coverage is below \(50\%\).

      \item[b] The dispersion index is recomputed after removing each
      region's single largest and three largest monthly withdrawal
      counts, and after excluding records with cluster-level or bulk-assigned withdrawal dates (the CAISO Cluster 15 block (\(303\) records) and all instances in which 100 or more records in a region share one withdrawal date (six dates, \(1\text{,}206\)). Overdispersion is attenuated but persists in every region (\(D \gg 1\)); the \(\chi^2\) goodness-of-fit test yields
      \(p < 0.001\) in all cases.

      \item[c] ``Excluding Imputed Dates'' excludes observations for
      which withdrawal dates were assigned using the date-imputation
      procedure.

      \item[d] Data: Rand et al.~(2026) \cite{rand2026}.
    \end{tablenotes}
  \end{threeparttable}
\end{table*}

We identified a total of \(39\) burst months across the seven ISO/RTOs between 2010 and 2025. The largest burst event occurred in CAISO in December 2024, coinciding with the deadline for its cluster study decision process. In that month, approximately \(168.6\) GW of capacity withdrew across \(426\) projects at a level of \(67.4\times\) the regional monthly mean. PJM also experienced its largest burst event in the same month (December 2024, \(43.4\) GW across \(485\) projects, \(18.1\times\) the mean). This event coincided with the deadline of transition queue processing of PJM. MISO, however, showed a different pattern. The largest withdrawals of MISO were distributed across four burst months in 2025 (July, September, October, and December) during the implementation of its reformed queue procedures. The December 2025 event was the largest, with \(42.0\) GW of capacity withdrawn at \(17.1\times\) the mean. ISO-NE (October 2025, \(31.4\times\) the mean) and SPP (November 2023, \(22.4\times\); August 2025, \(20.1\times\) the mean) also show withdrawal peaks coinciding with specific procedural milestones. 

\begin{figure*}[!t]
  \centering
  \includegraphics[width=6.5in]{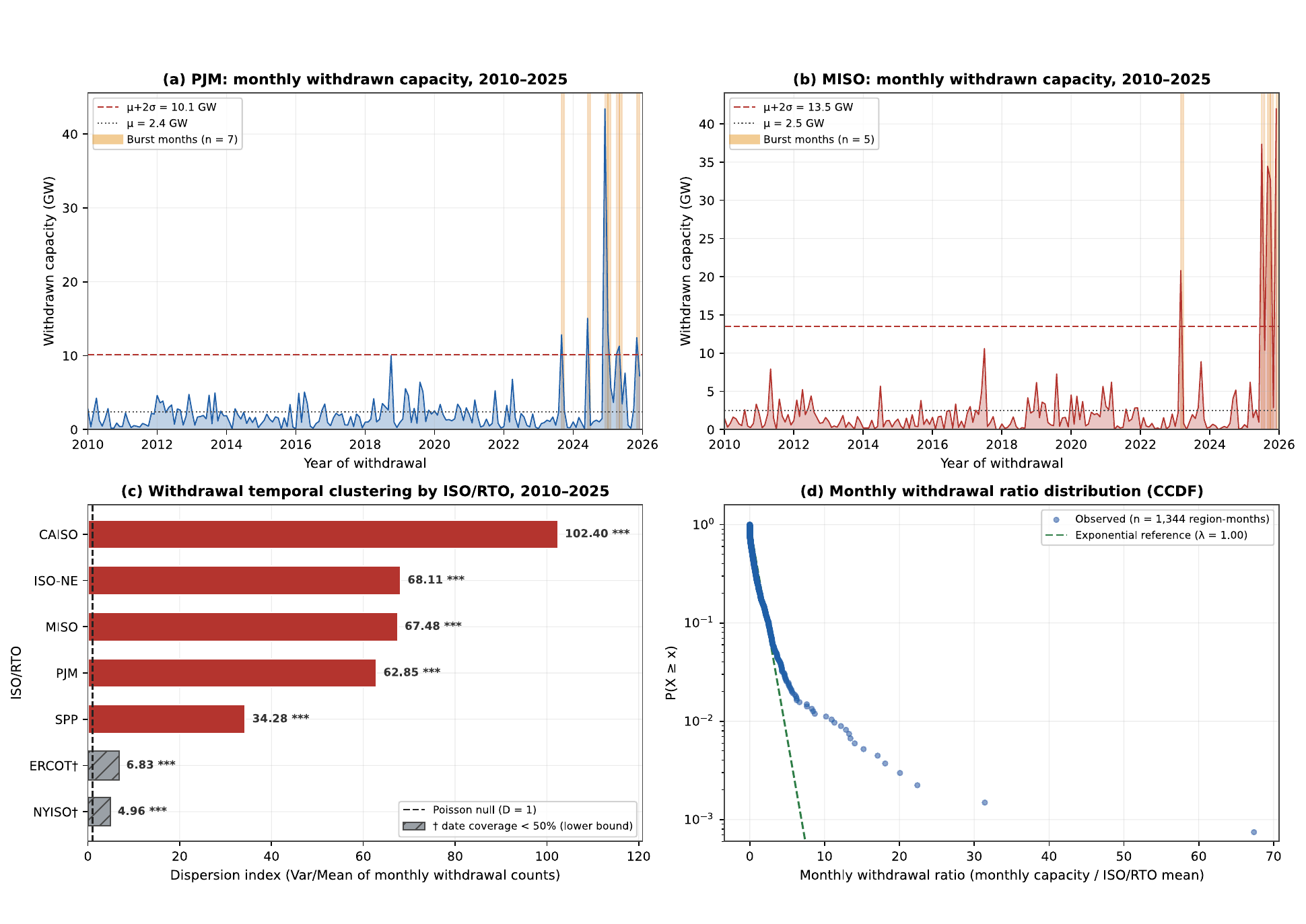}
  \caption{Empirical study results. (a) Monthly withdrawn capacity in PJM from 2010 to 2025, highlighting burst months and the \(\mu+2\sigma\) threshold. (b) Monthly withdrawn capacity in MISO. (c) Dispersion indices for each ISO/RTO, together with the Poisson reference line (\(D=1\)). (d) CCDF of the monthly withdrawal ratio (monthly withdrawn capacity normalized to each ISO/RTO mean, region-months pooled, \(n = 1\text{,}344\)), shown against a moment-matched exponential reference (\(\lambda = 1.00\)).}
  \label{fig:figure_2}
\end{figure*}

The CCDF of monthly withdrawal ratio deviates substantially from an exponential distribution in the upper tail (Fig.~\ref{fig:figure_2}(d)), consistent with the ''heavy-tailed'' behavior observed in cascading failures in other complex systems \cite{carreras2002critical, dobson2007complex}. The CCDF analysis pools normalized withdrawal ratios across seven ISO/RTOs that should be interpreted with caution. The distributional shapes differ between regions (regional skewness coefficients of \(1.9\text{--}13.1\); tail mass above five times the regional mean ranging from \(0.5\%\) to \(3.6\%\) of months), subject to differences in regulatory environments and study processes. The observed heavy-tailed behavior may then partly arise from heterogeneity in the underlying distributions across individual ISO/RTOs. Therefore, further analysis is required to stratify the dataset at ISO/RTO level, and to isolate endogenous dynamics of cascading withdrawals from heterogeneity inherent in the aggregated data. 

For each ISO/RTO, we calculated the conditional probability of co-withdrawal and compared it with the permutation null. In this case, we evaluated the intra-cohort correlation of withdrawal timing within the cohort peer sets. Under the unrestricted null, all seven ISO/RTOs show large excess co-withdrawal (\(2.5\text{--}10.0\) percentage points, \(5.1\text{--}10.3\) standard deviations). Under the entry-year-stratified null, which preserves the entry-to-exit ordering, the excess decreases to \(0.2\text{--}1.8\) percentage points. With null co-withdrawal bounded well above zero, percentage point differences understate the effect; expressed as a share of the range available above the null, the observed rates (\(88.1\text{--}97.5\%\)) close \(41\text{--}69\%\) of that distance. In the latter, the excess co-withdrawal is significant at the 5\% level in only four of seven regions: PJM (\(0.5\) pp, \(z = 2.74\), \(p = 0.003\)), SPP (\(1.0\) pp, \(z = 2.67\), \(p = 0.004\)), ERCOT (\(1.8\) pp, \(z = 2.52\), \(p = 0.009\)), and ISO-NE (\(1.5\) pp, \(z = 2.15\), \(p = 0.021\)). MISO (\(z = 0.89\), \(p = 0.21\)), NYISO (\(z = 1.28\), \(p = 0.12\)), and CAISO (\(z = 0.43\), \(p = 0.37\)) are not significant. 

With fixed entry and withdrawal dates, we permute the technology label to determine whether co-withdrawal concentrates within technology categories, independently of when projects entered or left. Under this test, the excess is significant in all seven regions: \(0.7\) percentage points for PJM (\(z = 2.99\)), \(0.9\) for MISO (\(z = 3.44\)), \(1.2\) for CAISO (\(z = 2.58\)), \(3.3\) for ERCOT (\(z = 3.69\)), 
\(3.4\) for ISO-NE (\(z = 3.90\)), \(2.1\) for SPP (\(z = 4.78\)), and \(5.0\) for NYISO (\(z = 4.23\)) (\(p \leq 0.005\)). The permutation results suggest concentration of withdrawal timing within technology categories in all seven ISO/RTOs. Co-withdrawal behavior within the same cohort, however, is significant in only four of seven regions and is small in magnitude. 

In addition to direct cost reallocation mechanisms, projects within the same cohort may be also jointly exposed to common shocks from study milestones, policy changes, and market conditions. The current testing framework does not separately isolate those shared exposures. By maintaining aggregate distribution of withdrawal dates for each region, the null model isolates temporal co-withdrawal correlations associated with specific cohorts and technology categories, and identifies temporal correlation associated with specific cohorts and technology categories. Distinguishing endogenous contagion from co-withdrawal responses to common exogenous shocks requires cost-allocation and restudy-trigger at project level, which are not currently available at the necessary level of granularity in the public domain. These findings are consistent with a contagion effect.

\subsection{Interpretation}
\label{subsec:interpretation}
\IEEEPARstart{T}{}he project withdrawals show statistically significant temporal clustering across all seven ISO/RTOs, with dispersion index ranging from \(5.0\) to \(102.4\) (all \(p < 0.001\)). The permutation tests further indicate co-withdrawal behavior among specific project cohorts and technology categories beyond overall regional clustering. These findings satisfy the necessary, though not sufficient, condition for the failure dynamics proposed in Section~\ref{subsec:cascading_failure_mechanisms}, indicating that the withdrawals do not occur as independent random events in the queue system. 

The largest withdrawal bursts often coincide with documented periods of queue stress, or procedural milestones associated with the regulatory feedback loop mechanism. However, the available data limits the determination of whether these events result from endogenous cascading effects, or from clustering due to a common underlying cause. However, the heavy-tailed distribution suggests that relatively small perturbations may trigger disproportionately large withdrawals, consistent with supralinear amplification of shocks in the queue system. This heavy-tailed distribution is also consistent with the network susceptibility to cascading failures. 

The results suggest that withdrawals are temporally clustered and dependent; however, they do not establish direct causal cascading events due to data limitations at project level. Nevertheless, the observed temporal distribution of project withdrawals and burst structure are consistent with the theoretical characteristics of cascading failure dynamics. Recent studies examine energy-only interconnection service directly \cite{norris2026energy, gorman2026review}, emphasizing that interconnection services do not guarantee deliverability, and that obligations associated with shared network upgrades are less stringent. 

Project withdrawals occurring after contractual commitment provide additional evidence on cost reallocation exposure consistent with cascading failure dynamics. Among interconnection requests that have executed interconnection agreements (IAs) signed between 2000 and 2022, \(34.8\%\) by project count and \(41\%\) by capacity had withdrawn by the end of 2025 \cite{rand2026}. They have already completed the required studies and established cost allocation arrangements involving contractual commitments. Therefore, the post-agreement withdrawals cannot be solely attributed to speculative applications. However, exposure to cost-reallocation and restudy risk after contractual commitment provides another possible explanation (Section~\ref{subsec:built_in_fragility_characteristics}). The amount of capacity exposed to such risks is substantial; as of end of 2025, \(549\) GW held a draft or executed IA without having reached commercial operation \cite{rand2026}.

\subsection{The 2025 Withdrawal Wave}
\label{subsec:the_2025_withdrawal_wave}
\begin{figure*}[!t]
    \centering
    \includegraphics[width=\textwidth]{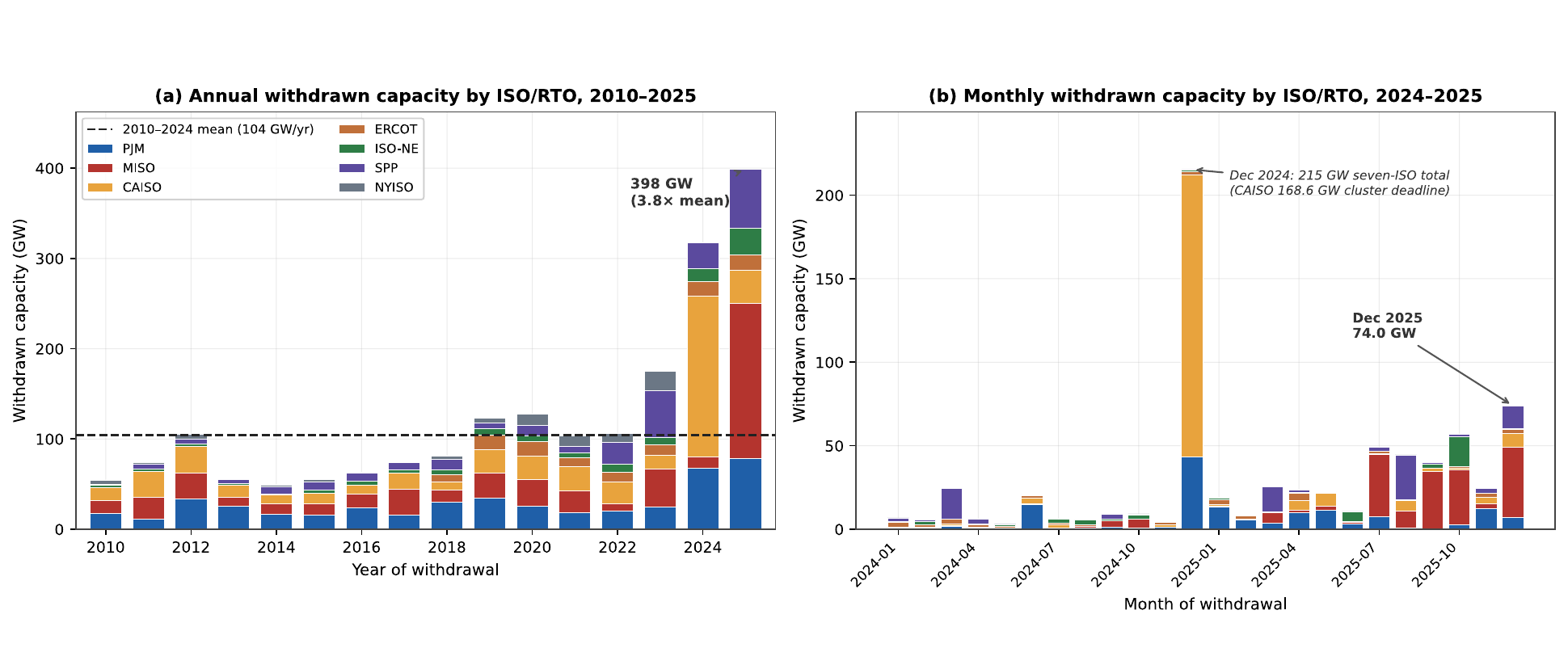}
    \caption{The 2025 withdrawal wave. 
    (a) Annual withdrawn capacity by ISO/RTO from 2010 to 2025. The \(2010\text{--}2024\) annual mean is shown for reference. (b) Monthly withdrawn capacity by ISO/RTO from January 2024 to December 2025; the December 2024 peak is the total of seven ISO/RTOs for that month. The cluster-deadline burst of CAISO contributes \(168.6\) GW (Section~\ref{subsec:empirical_analysis_results}). Only withdrawal records with timestamps are included (\(86.3\%\) of withdrawals across the seven ISO/RTOs). Data: Rand et al.~(2026)\cite{rand2026}.}
    \label{fig:figure_4}
\end{figure*}

\IEEEPARstart{T}{}he proposed framework predicts that project withdrawals temporally cluster around reform-related decision deadlines, particularly when projects receive updated information about costs and readiness requirements within similar times. The 2025 data provide an out-of-sample test of this prediction. Time-stamped records from seven ISO/RTOs account for a total of \(2\text{,}417\) project withdrawals, or roughly \(398\) GW of withdrawn capacity (Fig~\ref{fig:figure_4}). Relative to the annual average over \(2010\text{--}2024\), the number of projects withdrawn in \(2025\) is approximately \(3.5\) times higher by project count and \(3.8\) times by capacity. By both project count and capacity, this was the largest withdrawal year in the dataset. 

These withdrawals exhibit a temporal and geographic clustering. MISO alone accounted for \(171.7\) GW of project withdrawals over the year, of which \(146.4\) GW occurred in four burst months. Total monthly withdrawals across the seven ISO/RTOs reached \(74.0\) GW in December 2025. This wave of withdrawals occurred during the first full implementation cycle of FERC Order 2023 cluster reforms. Readiness-deposit requirements and cluster decision deadlines particularly compelled many projects to reassess their continued participation at critical withdrawal decision points. 

We establish two complementary interpretations for the 2025 withdrawal wave. First, the screening effect introduced by the FERC Order 2023 reforms purged speculative entries from the interconnection queue. As readiness requirements took effect, projects with limited development viability, or projects occupying queue positions for purely speculative purposes, withdrew from the interconnection queue. Under the second interpretation, the regulatory feedback mechanisms described in Section~\ref{subsec:cascading_failure_mechanisms} also contributed to the withdrawal clustering. Because the reforms impose standardized reform deadlines, individual withdrawal decisions can be temporally synchronized, making them a correlated shock at the system level. 

These mechanisms may operate simultaneously. The propagation component of the regulatory feedback mechanism depends on restudy cycles, though, and those effects can only be evaluated over subsequent data vintages. The 2025 data show that the system is exposed to the risk of synchronized mass withdrawals. Roughly \(25\%\) of the active projects waiting for access at the seven ISO/RTOs withdrew within a single year, which is not previously observed in the dataset. The magnitude of this coordinated withdrawal corresponds to the calibrated scenario in the computational model (Section~\ref{subsec:computational_contagion_model_results}, Experiment 7), which yields \(38.5\%\) total failure at moderate connectivity (\(k = 10\)).




\section{Computational Contagion Model of Queue Fragility}
\label{sec:computational_contagion_model_of_queue_fragility}

\subsection{Model Description}
\label{subsec:model_description}
\IEEEPARstart{W}{}e examine whether the structural properties of the queue system identified in Sections~\ref{sec:the_interconnection_queue_as_a_complex_adaptive_system} and~\ref{sec:preliminary_empirical_analysis} are sufficient to generate the cascading failure dynamics hypothesized in Section~\ref{subsec:formal_mapping_to_cascading_failure_models}. The computational model adapts elements of the interdependent-network percolation mechanism proposed by Buldyrev \etal ~(2010) \cite{buldyrev2010catastrophic} and the threshold models of cascades introduced by Watts (2002) \cite{watts2002simple} to the queue. 

A queue study cluster is represented as an undirected graph \(G = (V,E)\). Each node \(v\in V\) corresponds to a queued project, and each edge \((u,v) \in E\) represents a cost-sharing interdependency between two projects. The network structure follows an Erd\H{o}s--R\'enyi random graph \(G(n,p)\), with \(p = k/(n-1)\), where \(k\) denotes average degree. We evaluate \(k \in \{3, 5, 10, 20\}\) for low to high levels of cost-sharing connectivity. 

Each project is assigned a base cost load \(L \sim U(0.3, 0.6)\) and viability threshold \(T \sim U(0.55, 0.95)\). In this way, we ensure \(P(L>T) \approx 1.0\%\) at initialization, and thus, pre-shock withdrawal failure is negligible. A project withdraws once its cumulative load exceeds the threshold. 

When an initial shock removes a fraction of nodes, their cost load is redistributed to active neighbors (nodes) connected to it in \(G\). Each edge represents a cost-sharing interdependency, with a redistribution factor of \(\alpha = 0.08\). Each active neighbor receives \(\alpha\) times the load of the failed node (not divided among neighbors); a design choice considered for the cost-sharing structure within the queue, where more total redistribution per failure arises from higher connectivity, and each project bears a share of upgrade costs that does not shrink as more projects share the infrastructure. This rule implies that the total load introduced into the network by a single failure \(k\alpha L\), so the ratio of introduced to removed load is \(k\alpha\); values above unity define a self-sustaining regime (Section~\ref{subsec:computational_contagion_model_results}). If the cumulative load of any neighboring node exceeds its viability threshold, that node withdraws. The resulting vacancy causes further redistribution across its adjacent nodes, and subsequently, the cost burden on the remaining neighboring projects increases. The cascading chain of failures terminates when no active node exceeds its threshold. 

To limit the severity of cascading effect, the circuit-breaker mechanism caps the cost redistribution per adjacent node at an absolute value \(\beta \in \{ 0.04, 0.03, 0.02\}\), so that each active neighbor receives \(min(\alpha L, \beta)\). Given \(\alpha = 0.08\) and \(L \sim U(0.3, 0.6)\), transfers range over \([0.024, 0.048]\) with mean \(0.036\). \(\beta = 0.04\) binds on \(33\%\) of transfers, \(\beta = 0.03\) on \(75\%\), and \(\beta = 0.02\) on all of them. At \(\beta = 0.02\), the cap is therefore equivalent to a uniform reduction in \(\alpha\) and is not a truncation of large transfers. The model parameter \(\beta\) is a stylized representation of circuit-breakers. In practice, the queue process may employ various forms of circuit-breaker mechanisms. It can suspend cascading restudies when the withdrawal rate exceeds a specified threshold. The process may cap the percentage of upgrade costs that may be reallocated during a single restudy cycle, or establish cost floors, below which any reallocated amounts would be absorbed by the transmission provider instead of being passed to remaining projects. The parameter \(\beta\) is an abstraction of these mechanisms, collectively representing a ''cost reallocation cap'' applicable to each pair of adjacent projects. Circuit-breaker mechanisms in financial markets, which temporarily halt trading during cascading sell-offs, provide the conceptual basis for this modeling choice. 

\subsection{Simulation Design}
\label{subsec:simulation_design}
\IEEEPARstart{F}{}or each parameter configuration, we conduct 200 Monte Carlo simulations, with each network consisting of \(n = 500\) nodes. Experiment 1 varies network connectivity \(k\), and initial shock fraction \(f \in \{ 0.01, 0.02, 0.05, 0.08, 0.10, 0.15, 0.20 \}\). Experiment 2 examines the concept of the circuit-breaker mechanism introduced in Section ~\ref{subsec:built_in_fragility_characteristics}. For this experiment, connectivity is fixed at \(k = 10\), and the upper threshold of the circuit-breaker \(\beta\) is varied according to the configurations described above. 

Experiment 3 tests the sensitivity of qualitative results to the assumed distribution of the financial viability threshold \(T\). In addition to the baseline specification, two alternative configurations are tested against the baseline \(T \sim U(0.55, 0.95)\). The first is a narrow distribution of \(T \sim U(0.65, 0.85)\), which eliminates all initialization failures but reduces threshold heterogeneity. The second is a shifted distribution of \(T \sim U(0.50, 0.90)\), which increases the degree of load-threshold overlap. All other parameters (\(n, \alpha\), \(k\), shock fractions) remain unchanged. 

In Experiment 4, the indivisible transfer rule (\(\alpha L\) per active neighbor) is then replaced by a divided alternative in order to directly evaluate the redistribution mechanism. Under the latter, \(\alpha L\) is divided evenly among all active neighbors of a withdrawn node. Experiment 5 uses the Barabási--Albert scale-free topology of the matching average network degree instead of the Erd\H{o}s--R\'enyi graph. Experiment 6 then compares targeted shocks that remove the highest-degree nodes against random shocks at \(k = 10\), capturing the concentration risk described in Section ~\ref{subsec:built_in_fragility_characteristics}.

Experiment 7 runs scenarios calibrated using empirical data, in which the shock fractions are set to burst severity values from the LBNL Queued Up 2026 Edition data. The scenarios correspond to median burst (\(3.4\%\) of the region's active queue), 75th percentile (\(8.0\%\)), the 2025 annual wave (\(24.9\%\)), and the CAISO December 2024 event (\(42.0\%\)). The node capacities are drawn from the empirical lognormal fit to the project sizes across the seven ISO/RTOs (\(\mu = 4.37, \sigma = 1.52\) in log-MW; median \(\approx 101\) MW). 

\subsection{Computational Contagion Model Results}
\label{subsec:computational_contagion_model_results}
\begin{figure*}[!t]
  \centering
  \includegraphics[width=6.5in]{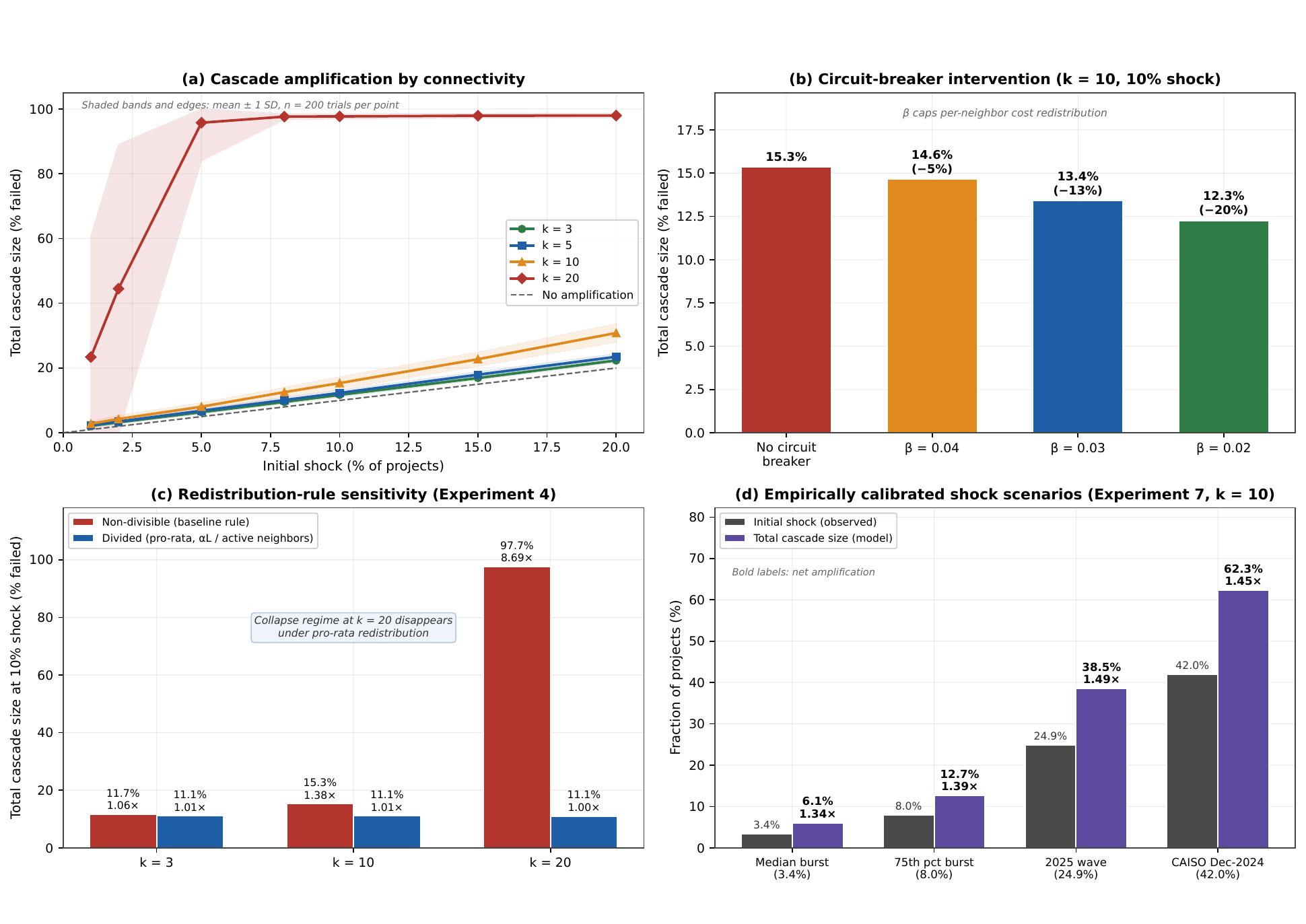}
  \caption{Computational model results. (a) Total magnitude of cascading failures varies as a function of initial shock under different levels of connectivity (\(k = 3, 5, 10, 20\)), along with a reference line representing the ''no amplification'' case. (b) Circuit-breaker intervention when \(k = 10\); the simulated severity of cascading failures is reduced by up to \(20\%\) through capping cost redistribution. (c) Redistribution-rule sensitivity (Experiment 4): under divided, pro-rata redistribution the collapse regime at \(k = 20\) disappears. (d) Shock scenarios (Experiment 7) with the corresponding shock magnitude and net amplification values.}
  \label{fig:figure_3}
\end{figure*}

\IEEEPARstart{A}{}cross all levels of connectivity, the total fraction of failures exceeds the initial shock fraction (Fig.~\ref{fig:figure_3}(a); Table~\ref{tab:table_4}). Under low connectivity (\(k = 3\)), a \(10\%\) initial shock results in a total failure of \(11.7\%\). Excluding the \(1.1\%\) zero-shock baseline, the net amplification is \(1.06\times\), implying a minimal cascade effect. At the moderate connectivity (\(k = 10\)), the same shock produces a \(15.3\%\) failure, corresponding to a \(1.38\times\) net amplification. In the model, interdependencies arising from cost-sharing cause failures to propagate beyond the initially affected projects. 

For network connectivity \(k = 20\), the queue system enters into a different operating regime. The mechanism is the undivided redistribution rule rather than graph density. At \(n = 500\), the \(k = 20\) network is still sparse (\(p = k/(n - 1) \approx 0.04\)). A withdrawing node transfers \(\alpha L\) to all its active neighbors, and thus, the total load introduced per failure becomes equal to \(k \alpha L\) against the failing node's own load \(L\). The ratio of introduced to removed load is \(k\alpha\), which is \(0.24\) at low connectivity (\(k = 3\)) and \(0.80\) at moderate connectivity (\(k = 10\)), but exceeds unity at high connectivity (\(k = 20\)) (\(k \alpha = 1.60\)). At this point, each failure adds more load to the system than it removes, and the cascade becomes self-sustaining. Even in the absence of an external shock, the zero-shock baseline itself becomes unstable (mean \(10.8\% \pm\) SD \(18.7\%\)) and yields a bimodal distribution clustering near two extremes of negligible failure or near-total collapse. Under external shocks, mean failure reaches near-total failure of \(95.8 \text{--}98.0\%\) across all tested shock levels. At this connectivity level even the \(\sim 1\%\) of nodes that initially satisfy \(L > T\) are sufficient to initiate full network failure in some simulation realizations. We therefore treat \(k = 20\) result as a boundary condition for network instability. 

\begin{table*}[!t]
  \renewcommand{\arraystretch}{1.3}
  \caption{Cascade amplification by network connectivity
  (\(n=500\), 200 trials)}
  \label{tab:table_4}
  \centering
  \begin{threeparttable}
    \begin{tabular}{ccccc}
      \toprule
      \textbf{Avg. Degree (\(k\))} &
      \textbf{5\% Shock \(\rightarrow\) \% Failed} &
      \textbf{10\% Shock \(\rightarrow\) \% Failed} &
      \textbf{20\% Shock \(\rightarrow\) \% Failed} &
      \textbf{Amplification (10\%)} \\
      \midrule

      \(k=3\)
      & 6.4\%
      & 11.7\%
      & 22.3\%
      & \(1.06\times\) \\

      \(k=5\)
      & 6.8\%
      & 12.3\%
      & 23.5\%
      & \(1.11\times\) \\

      \(k=10\)
      & 8.1\%
      & 15.3\%
      & 30.8\%
      & \(1.38\times\) \\

      \(k=20\)\tnote{a}
      & 95.8\%
      & 97.7\%
      & 98.0\%
      & \(8.69\times\)\tnote{b} \\

      \bottomrule
    \end{tabular}

    \begin{tablenotes}[flushleft]\footnotesize
      \item[a] The zero-shock baseline at \(k=20\) is bimodal and
      unstable across reseeding (mean \(10.8\%\) and SD \(18.7\%\)).

      \item[b] Amplification is defined at a \(10\%\) shock as \(A=(total failed - zero-shock baseline)/shock.\)
      Because of the unstable \(k=20\) baseline, the reported
      amplification is imprecise; recomputation using the updated
      baseline yields \(8.69\times\) rather than \(9.13\times\).

      \item[c] Simulation parameters are \(n=500\),
      \(\alpha=0.08\), \(L\sim\mathcal{U}(0.3,0.6)\),
      \(T\sim\mathcal{U}(0.55,0.95)\), and
      \(P(L>T\mid t=0)\approx1\%\), with 200 trials.

      \item[d] Zero-shock baselines are \(1.1\%\), \(1.2\%\), \(1.5\%\), and
      \(10.8\%\) for \(k=3,5,10,\) and \(20\), respectively.

      \item[e] Amplification is not scale-invariant. At \(k=20\),
      the same configuration yields \(17.0\times\) amplification
      under a \(5\%\) shock and \(4.6\times\) under a \(20\%\) shock.
    \end{tablenotes}
  \end{threeparttable}
\end{table*}

The transition between \(k = 10\) (stable, \(1.38\times\) net amplification under roughly \(10\%\) shock) and \(k = 20\) (unstable, bimodal collapse) suggests the existence of a critical connectivity threshold. Beyond this threshold, the system transitions from progressive failure, in which shocks produce proportionate additional failures, to a state of categorical fragility. In the latter, virtually any perturbation can trigger systemic collapse. Future research is required to evaluate intermediate connectivity levels (\(k = 12, 15\)) to locate the phase transition point more precisely. One resilience engineering intervention is to reduce the density of cost-sharing interdependencies, limiting the cascade risk.

Figure~\ref{fig:figure_3}(b) shows the effect of the circuit-breaker intervention at \(k = 10\) evaluated by Experiment 2. For a \(10\%\) shock, imposing circuit-breaker cap decreases the total failure from \(15.3\%\) (no intervention) to \(14.6\%\) (\(\beta = 0.04, 5\%\) reduction), \(13.4\%\) (\(\beta = 0.03, 13\%\) reduction), and \(12.3\%\) (\(\beta = 0.02, 20\%\) reduction). The corresponding reductions in mean per-neighbor transfer are \(3.7\%\), \(18.7\%\), and \(44\%\). The relationship is sublinear at the strictest setting, where a \(44\%\) reduction in the cost-transfer coefficient yields a \(20\%\) reduction in cascade size. Under the strictest cap constraint, net amplification decreases from \(1.38\times\) to \(1.08\times\). Assuming that the system experiences a connectivity-dependent phase transition, the results of these intervention suggest when the system operates near but below the phase-transition threshold, cascading failure propagation can be suppressed either by limiting cost redistribution per neighboring nodes, or by imposing a cap constraint that mimics a circuit-breaker mechanism. 

In Experiment 3, we examined the sensitivity of the model to the threshold distribution. A narrow distribution was used to reduce both the threshold heterogeneity and initialization failure rate (\(P(L > T) = 0\%\); \(T \sim U(0.65, 0.85)\)). The shifted configuration increased the initialization failure rate (\(P(L > T) \approx 4\%\); \(T \sim U(0.50, 0.90)\)). A phase transition from moderate to high connectivity appears under all configurations. When \(k = 10\), the net amplification is \(1.37\times\), \(1.05\times\), and \(1.97\times\) respectively under baseline, narrow, and shifted thresholds. At \(k = 20\), the corresponding mean failure is \(97.6\%\), \(82.0\%\), and \(98.9\%\) respectively. These results indicate that, at moderate to high levels of connectivity, the phase transition is not sensitive to the choice of threshold distributions tested in the study; however, the amplification effect and the location of phase transition still depends on the specific parameter values. 

Experiment 4 examined whether this instability depends on the redistribution rule. Under divided redistribution regime, as connectivity increases, the risk exposure faced by any individual project decreases. At \(k = 20\), the total failure decreases from \(97.7\%\) (non-divisible) to \(11.1\%\) (divided), and net amplification from \(8.69\times\) to \(1.00\times\). Under divided cost exposure, the phase transition is eliminated. The systemic instability exhibited by the model depends on whether the cost exposure of an individual project remains undiluted as the number of projects sharing the same infrastructure changes. 

Filings by grid operators document that withdrawals of peer projects subsequently increase upgrade obligations borne by the remaining projects \cite{miso2022, spp2021}. The cost differential between withdrawn and completed projects observed by Gorman \etal ~(2025) \cite{gorman2025grid} points to non-dilutive cost exposure. Neither rule is calibrated to observed cost allocation. Under the non-divisible rule, \(\alpha L\) is assigned to each active neighbor per failure, and the divided rule allocates it in total. The comparison therefore does not quantify any specific tariff reform and merely separates the structural driver. 

Experiments 5 and 6 altered topology and shock targeting. The scale-free network yields a greater failure at moderate connectivity (\(k = 10\)), \(26.7\%\) total failure versus \(15.3\%\) under Erdős--Rényi. At \(k = 20\), however, the collapse observed under the Erd\H{o}s--R\'enyi  topology does not occur (\(83.6\%\) versus \(97.7\%\)). The result points to the clustering of interdependencies around high-degree nodes. Failures propagate efficiently through nodes, while some lower-degree peripheral projects face comparatively limited exposure. At \(k = 10\), removing the highest-degree nodes increases total failure to \(17.6\%\), compared with only \(15.3\%\) under random shocks of the same size. 

Finally, Experiment 7 empirically evaluated the shock scale. Using the LBNL data, the model yields \(6.1\%\) total failure for the median burst (\(3.4\%\) shock; \(1.34\times\) amplification), \(12.7\%\) for the 75th-percentile burst (\(8.0\%\); \(1.39\times\)), \(38.5\%\) for the 2025 annual wave (\(24.9\%\); \(1.49\times\)), and \(62.3\%\) for the CAISO December 2024 event (\(42.0\%\); \(1.45\times\)). The net amplification remains stable within a range of \(1.34\times\) to \(1.49\times\) across more than an order of magnitude of empirically observed shock sizes. Under the assumed moderate connectivity (k = 10), the empirically calibrated shocks remain in the amplification regime and do not produce boundary-condition collapse. 

\subsection{Model Limitations and Validation}
\label{subsec:model_limitation_and_validation}
\IEEEPARstart{T}{}he model is deliberately stylized using a random graph topology rather than the actual interdependency structure of the interconnection queue system. The model does not fully reproduce entire institutional interdependencies of the queue. Given the complexity of the proportional-impact methods required by FERC Order 2023, we use a uniform cost reallocation rule in the model. The model relies on fixed thresholds and assumes a redistribution factor \(\alpha = 0.08\). The connectivity transition is governed by \(k\alpha\) and scales approximately \(1/\alpha\), and subsequently, the amplification magnitudes reported in Table~\ref{tab:table_4} are conditional on the \(\alpha\) value. However, developers typically make dynamic decisions in response to evolving costs, project timelines, and other project-specific conditions. The redistribution rule is non-conservative, in which each failure adds \(k\alpha\) obligation at the system level. This preserves individual exposure, but overstates actual cost reallocation, which conserves total upgrade cost. The collapse regime subsequently represents stylized accounting and does not predict tariff behavior.

The phase transition does not occur under fully pro-rata redistribution (Experiment 4), and it depends on whether individual projects must absorb undiluted reallocated costs. Under mass-conserving pro-rata redistribution, total failure peaks at moderate connectivity and then declines, so the monotone trend reported here is specific to the non-dilutive rule. The systemic collapse regime predicted by the model is associated with the empirically documented institutional property of cost allocation. Allocation rules that are designed to reduce or cap the cost exposure of individual projects, therefore, offer a directly testable design for suppressing the collapse regime. We refer to such rules as ''cost-isolation mechanisms''. Examples include pro-rata allocation and caps on per-project exposure. 

We aim to establish a proof-of-concept that the architectural properties of the queue system are sufficient to induce amplification effect and phase transition behavior, and that introducing circuit-breaker interventions can improve the system-level outcomes. Infrastructure interdependency modeling \cite{ouyang2014review} and resilience quantification \cite{bruneau2003framework, panteli2015influence} provide foundations for the higher-fidelity models in future work.

The amplification magnitudes shown in Table~\ref{tab:table_4} depend on the specification of parameter values (\(\alpha\), load and threshold distributions, graph topology). The phase transition occurs under all tested threshold distributions in the model; however, its location depends on the model parameters and shifts from the \(k = 10\text{--}20\) region under the baseline to the \(k = 5\text{--}10\) region under shifted thresholds. The amplification magnitudes also vary by a factor of \(4\text{--}6\times\) across configurations. Detailed empirical validation, however, requires project-level cost-allocation records and interconnection-study milestone timestamps. These data currently remain fragmented across Open Access Same-Time Information System (OASIS) portals of individual ISO/RTOs and have not yet been consolidated into a standardized, research-ready dataset.




\section{Policy and National Security Dimensions}
\label{sec:policy_and_national_security_dimensions}

\subsection{Resilience Standards for the Interconnection Queue}
\label{subsec:resilience_standards_for_the_interconnection_queue}
\IEEEPARstart{U}{}nder National Security Memorandum 22 (NSM-22), which superseded Presidential Policy Directive 21 in April 2024, the energy sector is already recognized as one of the sixteen designated critical infrastructure sectors. This designation is also stipulated in the frameworks administered by the Cybersecurity and Infrastructure Security Agency (CISA). The U.S. interconnection queue shows a high degree of interdependence, which is evident in shared network upgrades (Section~\ref{subsec:built_in_fragility_characteristics}), reallocation of upgrade costs to remaining projects (Experiment 4), and statistically significant temporal clustering of project withdrawals across all seven ISO/RTOs (dispersion index \(D = 5.0\text{--}102.4\), all \(p < 0.001\); Table~\ref{tab:table_2}). Although the queue system can be classified as critical infrastructure under NSM-22 and CISA, it has not yet been subject to a structured resilience analysis as developed in Sections~\ref{sec:preliminary_empirical_analysis} and~\ref{sec:computational_contagion_model_of_queue_fragility}.

Four policy actions are therefore recommended. First, transmission providers should model and report the dispersion and cascade-amplification metrics defined in Sections~\ref{sec:preliminary_empirical_analysis} and~\ref{sec:computational_contagion_model_of_queue_fragility} for their active interconnection queue portfolios. A systemic risk assessment of interconnection queue should be established, modeled structurally on the Dodd-Frank Act financial stress tests. Second, queue operators should publish standardized cascading risk metrics (Sections~\ref{sec:preliminary_empirical_analysis} and~\ref{sec:computational_contagion_model_of_queue_fragility}). Such transparency measures would allow developers and regulators to assess cascade vulnerability and cost redistribution exposure across the queue. Third, NERC or FERC should establish an interconnection resilience standards body, or jointly administered body, responsible for developing and enforcing minimum resilience standards for interconnection architecture. Fourth, research on interconnection system resilience can be supported through the DOE grid modernization portfolio, the National Science Foundation, or ARPA-E. 

\subsection{Implications for AI Data Center Infrastructure}
\label{subsec:implications_for_ai_data_center_infrastructure}
\IEEEPARstart{P}{}lanned hyperscale AI data centers require a power supply of \(100\) MW\(\text{--}1\) GW per site. The development pipeline capacity already exceeds \(100\) GW nationally \cite{doe2025b}. Each facility enters the interconnection queue system directly via collocated generation, or indirectly through contracted generation resources to serve that facility. Under the DOE Large Load Interconnection Directive \cite{doe2025a}, a new load above \(20\) MW is subject to federal jurisdiction. Subject to FERC June 2026 show cause orders, six major ISO/RTOs are to justify or reform their large-load tariff provisions \cite{ferc2026a}. Further, the Ratepayer Protection Act (H.R. 9340) would amend Section 111(d) of the Public Utility Regulatory Policies Act to direct state regulatory authorities to consider adopting a large-load cost standard for facilities above \(100\) MW of peak demand \cite{house2026}. The October 2025 directive addresses loads above \(20\) MW. The show cause orders define large loads as those exceeding \(50\) MW of peak demand interconnecting above \(69\) kV, and H.R. 9340 sets its threshold at \(100\) MW \cite{doe2025a, ferc2026a, house2026}. Regardless of the pathway used to enter the system, data center schedules remain exposed to cascading restudy and cost-reallocation mechanisms documented in the present paper. Queue system fragility, therefore, represents a structural constraint on the deployment of AI data centers. 

Accelerating data center interconnections without first addressing the queue system fragility would only shift the risk of cascading failure elsewhere in the system. The large load priority lanes proposed by FERC divert study resources away from the standard queue. Delays or withdrawals that arise from higher attrition among standard queue applicants, particularly contracted renewable generation projects planned to support data center sustainability commitments, could slow overall power sector decarbonization. Our results suggest that scalable grid expansion is constrained by fragility of the queue system. 

\subsection{Comparative Institutional Architecture}
\label{subsec:comparative_institutional_architecture}
\IEEEPARstart{T}{}he cascading failure mechanism identified in the present study belongs to a particular class of queue architecture, where applicants are serially processed project by project and shared upgrade costs are allocated among multiple projects. This creates coupling between independent applications. A single withdrawal initiates mandatory restudies and reallocates upgrade costs to other projects in the network. For withdrawals to spread across the system, this interdependence must be present, and it is what makes ''withdrawal contagion'' occur. 

Alternative architectures can eliminate cost-sharing interdependencies entirely. Integrated Resource Planning (IRP), centralized generation siting, and coordinated transmission-and-generation planning, for instance, do not allocate shared upgrade costs on a project-by-project basis. Consequently, the cost-reallocation interdependency and the failure contagion mechanism are eliminated. 

Each architecture, however, involves its own trade-offs; for example, between flexibility and coordination, market efficiency and planning efficiency, and developer autonomy and system optimization. By this analysis, queue fragility is not an inherent cost of market-based electricity systems. Rather, it is a consequence of the way most ISO/RTOs currently process interconnection requests. Eliminating shared-cost interdependencies across interconnection applications can suppress the identified cascading failure, without impacting the competitive electricity market design.



\section{Conclusion}
\label{sec:conclusion}
\IEEEPARstart{T}{}he U.S. interconnection queue system was designed for an environment characterized by a relatively lower volume and lower complexity. Today, however, it exhibits the ''cascading withdrawal failure'' dynamics described in Sections~\ref{sec:preliminary_empirical_analysis} and~\ref{sec:computational_contagion_model_of_queue_fragility}. Current reforms may mitigate the application backlog by accelerating study throughput. However, the cascading failure mechanisms remain unaddressed at the system architecture level. 

The statistical analysis reports that project withdrawals are temporally clustered across all seven ISO/RTOs, with dispersion indices ranging from \(5.0\) to \(102.4\) (\(p < 0.001\)). By the empirical analysis, we further identify 39 monthly withdrawal bursts, where withdrawn capacity reached as high as \(67.4\times\) the regional mean. Permutation tests show that withdrawal timing concentrates within technology classes in every region. Co-withdrawal within cohorts is significant in four of seven regions and small under fixed entry-exit trend. The withdrawal series are therefore not independent draws, but the association established is temporal and technology specific. Under low to moderate connectivity (average degree \(k = 3\text{--}10\)), the stylized Erd\H{o}s--R\'enyi network model shows that tight interdependence, undiluted cost reallocation, and heterogeneity in project thresholds are sufficient, in the model, to yield net cascade amplification in the sub-doubling range (\(1.0\text{--}1.5\times\)). Once the critical connectivity threshold is exceeded, the simulated system transitions to a regime of near-total failure. This threshold is located by the product \(k\alpha\) and it does not arise under mass-conserving pro-rata redistribution (Section~\ref{subsec:model_limitation_and_validation}). Under moderate connectivity (\(k = 10\)), the simulated circuit-breaker mechanisms mitigate the magnitude of cascading failures in the model by up to \(20\%\) relative to the no-intervention baseline. 

Since the interconnection queue is the primary gateway to access the grid for all new generation and storage in the United States, the study completion and withdrawal rates of the queue system directly affect the pace of grid decarbonization. The queue system also influences the ability to secure reliable power supply at predictable costs for critical infrastructure, particularly data center and defense-sector loads. The resulting delayed capacity implies that the power generation capacity originally scheduled to come online under the energy sector transition plan was not brought into operation as planned, and foregone emissions reductions and economic value, which we do not quantify in the present study. 

The queue system has become a major structural bottleneck between the existing clean energy project pipeline and a decarbonized electricity system. The results offer both the empirical and computational foundations for resilience engineering research of the queue system and cost-isolation mechanisms. 

Data Availability. The LBNL Queued Up dataset is publicly available at https://emp.lbl.gov/queues. The dataset used is the public LBNL data file through 2025 \cite{rand2026}. The empirical analysis code, the computation contagion model, and the figure generation scripts are available from the corresponding author on request.



\bibliographystyle{IEEEtran}
\bibliography{references}

\end{document}